\documentclass[12pt, draftclsnofoot, onecolumn]{IEEEtran}
\usepackage{graphicx}
\usepackage{xcolor}
\usepackage{multicol}
\usepackage{amssymb}
\usepackage{cite}
\usepackage{amsmath}
\usepackage{amsfonts}
\usepackage{amsthm}
\usepackage{subfigure}

\def\bw{\mathbf{w}}
\def\bh{\mathbf{h}}
\def\bq{\mathbf{q}}

\def\twth{\tilde{\bw}_\theta}
\def\C{\mathbb{C}}
\def\R{\mathbb{R}}

\def\mb{\mathbf}

\usepackage{algorithm}
\usepackage{algorithmic}
\usepackage{comment}
\newcommand{\cmnt}[1]{\hspace*{-1px}}
\usepackage{tikz}
\usepackage{epstopdf}
\usetikzlibrary{calc,shapes.geometric}
\DeclareMathOperator*{\dg}{diag}
\title{Energy-preserving Indirect-feedback for Wireless Power Transfer}
\author{Siddhartha~Sarma, Rahul~Shrestha,~\IEEEmembership{Senior Member,~IEEE}, and Rohit~B.~Chaurasiya,~\IEEEmembership{Member,~IEEE}
\thanks{This work has been submitted to the IEEE for possible publication. Copyright may be transferred without notice, after which this version may no longer be accessible.}
\thanks{S. Sarma and R. Shrestha are with the School
	of Computing and Electrical Engineering, Indian Institute of Technology Mandi,
	Himachal Pradesh, 175005, India (e-mail: \{siddhartha, rahul\_shrestha\}@iitmandi.ac.in). R. B. Chaurasiya is with International Institute of Information Technology Naya Raipur, Chhattisgarh, 493661, India (e-mail: rohitc@iiitnr.edu.in)}}
\begin{document}
	\maketitle
	\begin{abstract}
		Recognising the limitations of various existing channel-estimation schemes for energy beamforming, we propose an energy-preserving indirect feedback-based approach for finding the optimal beamforming vector. Upon elaborating on the key ideas behind the proposed approach---dynamics of the harvest-then-transmit protocol and the latency associated with the charging process---we present an algorithm and its hardware architecture to concretise the proposed approach. The algorithm and the hardware architecture are supplemented by mathematical analysis, numerical simulation and hardware utilisation details, ASIC synthesis and post-layout simulation details, respectively. We firmly believe this paper, due to its unified algorithm-hardware design, will open up new avenues for research in radio frequency (RF) wireless power transfer.
		
	\end{abstract}
 \begin{IEEEkeywords}
 	Wireless power transfer, Energy beamforming, MISO system, Indirect feedback, Harvest-then-transmit, Charging circuit, Hardware architecture, ASIC
 \end{IEEEkeywords}
  \section{Introduction}
  Wireless power transfer (WPT) through radio frequency (RF) signals has been envisaged as one of the key drivers for the mass adoption of the internet-of-things (IoT). The benefits of the WPT over the conventional power delivery method, i.e., employing batteries,  are several, such as reduction in the weight of devices, curtailment of the maintenance cost, and relief from the tiresome routine of replacing or recharging on-device batteries. Additionally, the WPT is eco-friendly as it curbs e-waste generation by alleviating the usage of batteries. The adoption of WPT will usher in the era of true mobility as both information and power will be delivered wirelessly. However, there are several technological challenges to transform the WPT from a promising technology to full-fledged solution for powering the next generation IoT systems. Hence, such challenges must be addressed foremost to proliferate the deployment of WPT.
  
  The WPT has been plagued by its low end-to-end energy transfer efficiency since its inception. Primarily, this is attributed to the omnidirectionality of the electromagnetic radiation from standard radiative elements, e.g., dipole antennas. To address this issue, \emph{energy beamforming} with the help of a multi-antenna transmitter has been reported in literature \cite{juTWC2014}, where the electromagnetic energy radiated by an antenna array is focused in a direction by adjusting gains and phases of individual antenna elements. However, to perform energy beamforming, a transmitter must estimate the channel between itself and a receiver by enrolling help from the receiver, leading to additional challenges, especially when the receiver is under stringent power constraints.
  %However, to perform energy beamforming, the channel state information (CSI) between a transmitter and a receiver must be available at transmitter. \txtblu{This} brings forth additional challenges, especially when receivers are working under stringent power constraints.
   
   The necessity of estimating downlink wireless channels is already well-motivated in the multi-antenna, single as well as multi-user wireless communication scenarios. As a result, various estimation schemes had been proposed in the wireless communication literature. Systems adopting time division duplexing (TDD) estimate the downlink channel with the help of uplink pilot tones by exploiting the channel reciprocity. On the other side, when frequency division duplexing (FDD) is employed, downlink pilots and the feedback from the receivers are essential for reliable estimates. % needed to estimate wireless channels reliably.

	It is apparent that both the aforementioned techniques require the active participation of receivers, and consequently, incorporating any of them into energy harvesting devices will cost a portion of the harvested energy. Subsequently, the depletion of harvested energy will lessen the throughput when the harvesting device transmits its information. The pilot-based schemes are also ill-suited for the scenarios where the information and energy transmission are scheduled in different frequency bands. \cmnt{The reason behind }This is because the reciprocity assumption is hardly valid in such cases, especially when those bands are far apart in the frequency spectrum or differ in \cmnt{terms of}bandwidth. Recognizing the limitation of the conventional channel estimation schemes concerning the wireless power transfer, we ask a question, \emph{``Is it possible to find the optimal beamforming vector for energy beamforming \cmnt{estimate the channel between a multi-antenna energy transmitter and an energy harvester} without costing any energy at the energy harvester?"}.

	We attempt to answer this question by delving inside the charging subsystem of a generic energy harvester device, often treated as a black box in the wireless communication literature. Furthermore, we exploit the temporal dependency of the information transmission on the energy harvesting process imposed by the prevalent \emph{harvest-then-transmit} protocol \cite{juTWC2014}. The dynamics of the harvest-then-transmit protocol, combined with the inherent delay present in the charging process, facilitate a unique way of obtaining feedback regarding the channel state information or a chosen beamforming vector. Yet, such unprecedented approach has not been adopted in the literature.
	
	%In the context of the WPT, a few contributions have \txtblu{Compared to various iterative single-bit or multi-bit feedback schemes proposed in the literature, the approach presented here will }
%	In comparison 
%	In the WPT literature, a few works proposed iterative schemes to optimize the (energy) beamforming vector by seeking single-bit or multi-bit feedback from the energy harvester at every iteration. Thus, it is absolutely clear that even with the most energy-efficient wireless transmission technology, such feedback-based schemes will cost at least a few micro Joules per packet to the harvesting device. In contrast, by adopting our indirect-feedback-based solution an energy harvesting device will be able to save this energy and utilise it for information processing or transmissions.
	
	 Compared to several iterative single or multi-bit feedback schemes \cite{xuTSP2014Onebit, chenTVT2013} proposed in the literature, the approach presented here is energy-preserving as it relies on indirect feedback. The energy saved in this manner can be utilised in information processing, reception or transmission at the energy harvesting device, resulting in the superior performance of energy harvesting systems.
	
%	\txtred{In this article, we first propose an indirect-feedback based multiple-input single-output (MISO) channel-estimation mechanism that enables a multi-antenna energy transmitter to estimate the channel between itself and an energy harvester without the requirement of any explicit participation from the energy harvester device for the channel estimation process (e.g., sending pilot signals or providing direct feedback to the energy transmitter). 

  For ease of reference, we list the major contribution of this article below. 
   \begin{itemize}
   	\item We first propose a scheme to find the optimal beamforming vector for a multiple-input single-output (MISO) wireless power transfer set up where no explicit participation, e.g., the transmission of pilot signals or direct feedback regarding the pilots transmitted by the energy transmitter, is needed at the energy receiver end. We also elaborate on the proposed scheme with the help of a schematic representation of the overall system.
   	
   	\item  We present an algorithm to find the optimal beamforming vector from the time-to-recharge values of probing vectors obtained from an orthonormal basis of $\C^N$, where $N$ is the number of antennas at the transmitter. By recognising the limitation of the proposed algorithm, we suggest modifications in the algorithm that results in curtailment of the probing duration (referred to as FAP phase in section \ref{sec:sysmod}).

%	With the help of a schematic representation of the overall system, we elaborate on the process of generating the optimal beamforming vector from the indirect feedback. } %\txtred{Furthermore, this work presents adequate explanation of various subtasks involved in the proposed channel-estimation mechanism. This paper also discusses a schematic representation of an overall system compliant with the proposed channel-estimation process for better clarity from the deployment aspect. Here, other subtasks of this system, except the suggested channel estimator, are standard designs used in practical wireless communication systems.}
	
	\item In order to facilitate the rapid deployment of our proposed solution in contemporary wireless devices/systems, this paper also present a new digital hardware architecture for the proposed optimal beamforming vector finding algorithm. % to obtain the optimal be for the channel estimator based on the proposed \txtblu{channel-estimation} algorithm. 
	To the best of the authors' knowledge, such a digital architecture is the first-time implementation in the WPT literature. Its micro-architecture design and hardware implementation results have been comprehensively presented here. The notion of such a hybrid design process is to shed ample information towards fine-tuning various design parameters of the proposed energy beamforming solution\cmnt{channel estimation} that leads to a superior and practical design\cmnt{that is} suitable for real-world applications. %Eventually, the proposed hardware architecture marks its design as the benchmark implementation for any similar prototypes in the future.
	
\end{itemize}
\subsubsection*{Organisation} Section \ref{sec:relwrk} lists various related works from where the current article has drawn inspiration and highlights the research gap addressed in this paper. In section \ref{sec:sysmod}, we elaborate on the system model considered in this paper. Two major contributions of this paper, namely, the algorithm to find the optimal beamforming vector and a digital hardware architecture for the same are presented in the sections \ref{sec:algo} and \ref{sec:hrdwr}, respectively. We present numerical results in section \ref{sec:numres} to demonstrate the performance of the proposed algorithm with respect to various system parameters. Lastly, we conclude the paper in the section \ref{sec:cnclsn} by reiterating and summarising the major contributions of this paper.
\subsubsection*{Notation} We denote vectors and matrices using lower-case and upper-case boldface fonts, respectively. $\mb{a}^\dagger$ represents complex conjugate transpose of a complex vector $\mb{a}$. Euclidean norm of a vector is denoted by $||.||$. $\C$ denotes the set of complex numbers, and $j$ is the complex number $\sqrt{-1}$. The magnitude, real and imaginary part of a complex number is denoted by $|.|$, $\Re\{.\}$ and $\Im\{.\}$, respectively. 

	\section{Related work}\label{sec:relwrk}
	
	As the early works on energy harvesting focused on establishing performance bounds, e.g., achievable throughput \cite{juTWC2014} and rate-energy trade-offs \cite{zhangTWC2013,leeTWC2015}, various issues pertaining to the actual implementations were either overlooked or superficially addressed. The predominantly adopted assumptions, such as availability of the channel state information, and linearity of the energy harvesting circuit, were relaxed in the later years, leading to a more realistic system model and refinement of the performance bounds presented earlier \cite{clerckxJSAC2019, boshkovskaTCOM2017}. The study of system models encompassing multi-antenna transmitters underwent a similar transition. While the focus of those studies remained the same---finding an optimal beamforming vector to optimize the utility under consideration---the complexity of resource allocation problems increased manifold due to the introduction of various practical constraints. In the next paragraph, we list a few notable works to highlight the importance of the availability of channel state information for optimizing throughput or transmit power or received power for multi-antenna WPT or simultaneous wireless information and power transfer (SWIPT) systems. In the subsequent paragraphs, we review different channel estimation schemes for multi-antenna WPT proposed in the literature and a few works that delve into the hardware architecture of their proposed strategies to position our work in the research landscape concerning wireless power transmission and harvesting. 
	
	Throughput maximization for multi-antenna wireless power transfer systems studied in \cite{LiuTC2014} was later revisited in \cite{boshkovskaTCOM2017} by relaxing the full CSI requirement and adopting a non-linear energy harvesting model. \cmnt{by considering a more realistic model} Maximizing the (weighted) sum received power at energy receivers was the objective of the studies carried out in \cite{xuTSP2014} and \cite{sonTWC2014}, with the latter extending the analysis to the imperfect CSI case. On the other hand, \cmnt{basing on the same assumptions mentioned earlier,} the authors in \cite{shiTWC2014} and \cite{timTWC2014} targeted the transmit power minimization problem for a single transmitter multi-user MISO SWIPT scenario and MISO interference channel with K transmitter-receiver pairs, respectively. The interference caused by signals intended for other receivers was mitigated by jointly optimizing the beamforming vector at transmitter(s) and power-splitting ratios at each mobile station.  
	
	Various approaches proposed in the literature for downlink channel estimation in WPT are nonetheless minor variations of their counterparts in wireless communication. Hence, many disregarded the differences between wireless communication and WPT systems. For example, WPT is unidirectional, unlike the data communication in wireless communications, and therefore, requires relatively simpler hardware for implementation than the latter. Additionally, if energy harvesting devices solely rely on the harvested energy for their operations, transmitting pilot or feedback signals for them could be burdening, if not practically realisable. Nevertheless, the schemes employing uplink pilot \cite{zengTCOM2014, zengTCOM2015} and received-power feedback \cite{yangTSP2014, leeTSP2016, choiTSP2016, choiTWC2016, xuTSP2016, abeywickramaTSP2018} provide frameworks for developing improved channel estimation schemes and benchmarking them.
	
	The transmitter side operations of our proposed scheme are similar to the received-power feedback-based schemes. However, the receiver side differs significantly in design as well as operations. In the received-power feedback schemes, a receiver measures the received power or energy, possibly with an energy meter \cite{xuTSP2014Onebit}, over an interval and feedback a quantized value of the measured power or difference of the measured power over two consecutive intervals to the transmitter \cite{xuTSP2016}. However, the quantization can be skipped if the transmitter refers to a pre-shared codebook for choosing a beamforming vector and the receiver feedback the code corresponding to the beamforming vector that resulted in the maximum received power \cite{chenTVT2013}. Two apparent shortcomings of the received-power feedback-based schemes are addressed in our proposed approach that exploits the dynamics of the harvest-then-transmit protocol. The first shortcoming is the requirement of an additional feedback module comprising an energy meter, quantizer, and additional RF components (if the feedback is sent over a different channel). The second one is the direct dependency of the channel estimation accuracy on the nature of feedback (partial/full) \cite{yangTSP2014} or the number of feedback bits \cite{xuTSP2016}. As the feedback considered in our work is indirect and the instrumentation for processing the feedback is part of the transmitter, the two shortcomings mentioned above are tackled.    
	
	The indirect feedback-based approaches, also termed blind or hidden feedback, appeared in wireless communication literature in the context of cognitive radio \cite{zhang2010, noam2013}.  In \cite{zhang2010}, the primary transmitter's power adaption acts as indirect feedback to the secondary users, which is employed to reduce the interference caused to the primary receivers by the secondary transmitters. A similar approach was proposed in \cite{noam2013} to find the null space of the MIMO channel between the secondary transmitter and the primary receiver for an underlay cognitive radio network. However, to the best of our knowledge, finding the optimal beamforming vector for WPT by employing indirect feedback has not appeared in the literature.
	
	Hardware implementation aspect of WPT systems has been sparsely reported in the literature. Recently, Samith et al. demonstrated the system-level hardware implementation in \cite{choiTWC2016, abeywickramaTSP2018} that enables energy beamforming by using received signal strength indicator (RSSI) value for the channel estimation. Here, the experimental validation has been carried out with the aid of USRPs and commercially available ultra-low power microcontrollers. In \cite{mishraTCSIIB2015}, hardware validation of wireless energy transfer to
	the onboard (P1110 evaluation board) energy storage element has been demonstrated with the aid of a dedicated RF energy source in conjunction with 6.1 dBi antenna. To the best of authors’ knowledge, dedicated digital hardware-architecture suitable for field-programmable gate-array (FPGA) and application-specific integrated-circuits (ASIC) implementation platforms for any indirect feedback-based energy beamforming is unexplored till date. Hence, by proposing a generic and new digital architecture for the same this work will play key role in the development of hardware- and energy-efficient WPT systems.  %deployment of WPT systems with better energy and hardware efficiencies for longer battery life and low cost, respectively. 
	Aforementioned holistic approach of designing a new algorithm and its corresponding architecture is an absolute necessity for transforming an unprecedented idea into real-world implementation. Such contributions are widely adopted by various state-of-the-art implementations \cite{chaurasiyaTCE2022,vermaTVT2022}.
	
%   \begin{figure}[!h]
%   	 \centering
%   	  \includegraphics[scale=1]{tikzFigures/conventionalEstimationProcesses}
%   	  \caption{Conventional downlink channel estimation mechanisms employed in wireless communication systems: (a) uplink pilots transmission from a mobile station, exploiting the channel reciprocity; (b) Collecting feedback from the mobile station after transmitting downlink pilots.}
%   	  \label{fig:cnvntnlChnnlEstm}
%   \end{figure}
\begin{figure}[!h]
	\centering
	\includegraphics[scale=0.6]{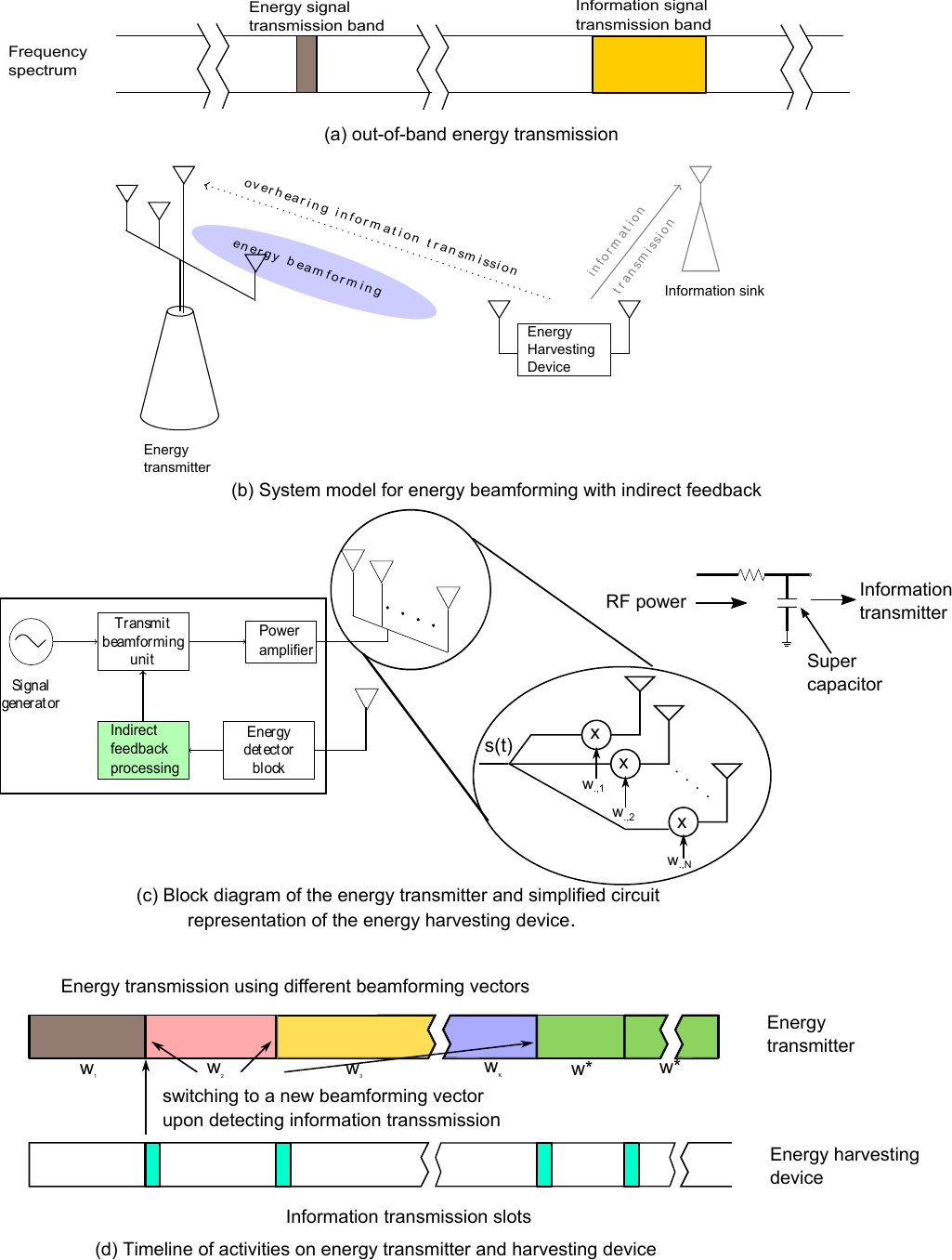}
	%      	\includegraphics[scale=0.6]{tikzFigures/energyBeamformerBlockDiagram}
	%      	\caption{(a) Channel estimation by using different beamforming vectors ($\mb{Q}$ is a unitary matrix, refer to the Algo. \ref{algo:blindChnlEst} ) in different slots, (b) block diagram of the energy transmitter}
	%      	\label{fig:timeSlotBlckDgrm}
	\caption{A system level overview of the indirect feedback based energy beamforming scheme.}
	\label{fig:systemOverview}
\end{figure}  
\section{System model}\label{sec:sysmod}
	The system model considered in this article is similar to the one mentioned in \cite{choiTWC2016, abeywickramaTSP2018, zhangArxiv2022}, where a multi-antenna energy transmitter transfers energy via RF signals to a single antenna energy harvester by employing energy beamforming. However, unlike \cite{abeywickramaTSP2018}, we assume an out-of-band separated receiver architecture, i.e., energy harvester and information transceiver have different antennas and RF blocks and they operate on different frequency bands \cite{LuCST15} (refer to Fig. \ref{fig:systemOverview}(a)). There are two major reasons for considering this architecture. First, decoupling the energy harvesting, and information transmission processes and the associated hardware becomes feasible---a prerequisite for our indirect feedback-based energy beamforming scheme. Second, it has been adopted in commercial products, e.g., Powercast energy harvesters \cite{powercast}, an evaluation board predominantly used in the literature to validate various theoretical findings.%Second, it has been adopted in widely \txtred{considered} commercial energy harvester, Powercast P2110 \cite{powercast}. 
	
	Specifically, in our system model, we assume that the energy transmitter has $N$ antennas supported by the associated hardware for energy transmission and one dedicated antenna backed by an information processing unit \cmnt{RF circuitry }for overhearing the (information) transmission of the energy harvester device.
	On the other hand, the energy harvester is equipped with two antennas and associated RF blocks--- one for energy reception and the other for communication. Additionally, we assume that the channel between the transmitter and receiver varies slowly due to their proximity and immobility. 
	
	The block diagram of our system model, shown in Fig. \ref{fig:systemOverview}(b), depicts an energy transmitter with two sets of antennas, an energy receiver (information transmitter) and an information sink along with the directions of energy transmission, information transmission and indirect feedback. In Fig. \ref{fig:systemOverview}(c), we present a schematic overview of the internal blocks of the energy transmitter. %The functioning of the indirect feedback processing \cmnt{channel-estimation} block is discussed later.
\subsection{Signals transmitted and received}
We assume that the energy transmitter generates a single tone signal, $s(t) = \sqrt{2}\cos(2\pi f_c t)$, having frequency $f_c$, and multiplies it with a unit-norm beamforming vector $\bw \in \C^N$. Following that, it transmits the signal through antennas after amplifying them by a set of power amplifiers whose amplification factor is governed by the transmit power ($P_t$) of the energy transmitter. Note that though we are considering a narrow-band (single tone) signal, the proposed algorithm can be applied to wideband signals, e.g., multi-tone signals \cite{huangTSP2017}, with little or no changes. 

We express the transmitted signal from the energy transmitter as
\begin{align}\label{eq:txSignal}
\mb{y}(t) = \sqrt{P_t}\bw s(t).
\end{align}
Assuming that the channel gain between the transmitter and the energy harvester is denoted by a complex vector $\bh \in \C^N$, we write the received signal at the energy harvester as 
\begin{align}
r(t) = \sqrt{GP_t}\bh^\dagger\bw s(t),
\end{align}
where the constant $G$ captures the effect of antenna gain, polarization, and wavelength of the transmitted signal, etc. on the received signal \cite{griffinAPM2009}. The average received power at the harvester is obtained by averaging the absolute square of the received signal over a sufficiently long interval and\cmnt{. Following the standard reduction techniques, we can express the received power as } can be expressed as
\begin{align}\label{eq:rxPower}
P_r = GP_t|\bh^\dagger\bw|^2.
\end{align}
From \eqref{eq:rxPower} it is apparent that for a given transmitter-receiver set up (i.e., for a fixed $G$) and transmit power, the received power will be maximized if the beamforming vector ($\bw$) is chosen in such a way that $|\bh^\dagger\bw|$ is maximized. As $\bw$ is assumed to be a unit-norm vector, it can be easily shown that $\bw = \bh e^{j\theta}/||\bh||$ with an arbitrary $\theta$ is the desired beamforming vector. 

\subsection{Energy harvester model}
%Now we focus on the energy harvester to see how it utilizes the received signal to harvest energy from it. The received signal is an RF single tone signal, possibly, distorted by the non-linearities present in the RF reception block of the energy harvester. This signal needs to be rectified and filtered before it can be used to charge a super-capacitor for energy storage. %The RF-to-DC conversion efficiency, also referred as harvesting efficiency, primarily depends on the characteristics of the diodes that are \txtred{integrated} in the rectifier circuit. 
%The RF-to-DC conversion efficiency, also referred to as harvesting efficiency primarily depends on the characteristics of the diodes and transistors in the rectifier circuit. While several early works on SWIPT and WPCN studied the performance of energy harvesting devices in various settings, e.g., multi-hop, multi-user networks, ignoring this important fact (cf. \cite{LuCST15} and references therein),
%%that focused on the performance analysis through mathematical models assumed the harvesting efficiency as constant, 
%recently, researchers have started to acknowledge the importance of incorporating the non-linearity of the efficiency function in their system models to have more accurate characterisation of the performance of energy harvesting systems \cite{clerckxJSAC2019}. 

Now we focus on the energy harvester and explain the harvesting process. The received signal is a single-tone RF signal, possibly, distorted by the non-linearities present in the RF reception block of the energy harvester. This signal is rectified and filtered to produce a DC signal suitable for charging a super-capacitor. 

The RF-to-DC conversion efficiency, also referred to as harvesting efficiency, is the ratio of the powers of the received RF and the output DC signals. As the rectifier circuit comprises various non-linear components, e.g., diode and transistors, the nonlinearity of the harvesting efficiency is evident. Therefore, an accurate characterisation of the performance of energy harvesting systems requires one to incorporate this non-linearity in system models, which was neglected in several early works. In this work, we assume that the harvesting efficiency belongs to a class of non-negative, non-linear functions defined over an interval of positive real numbers. 
%In this work, we assume that the harvesting efficiency belongs to a class of non-negative non-linear functions defined over the positive real values and satisfies an \txtred{easily justifiable} constraint. 

\subsubsection*{Non-linear harvesting efficiency}
\begin{figure}[!h]
	\centering
	\includegraphics[scale=0.45]{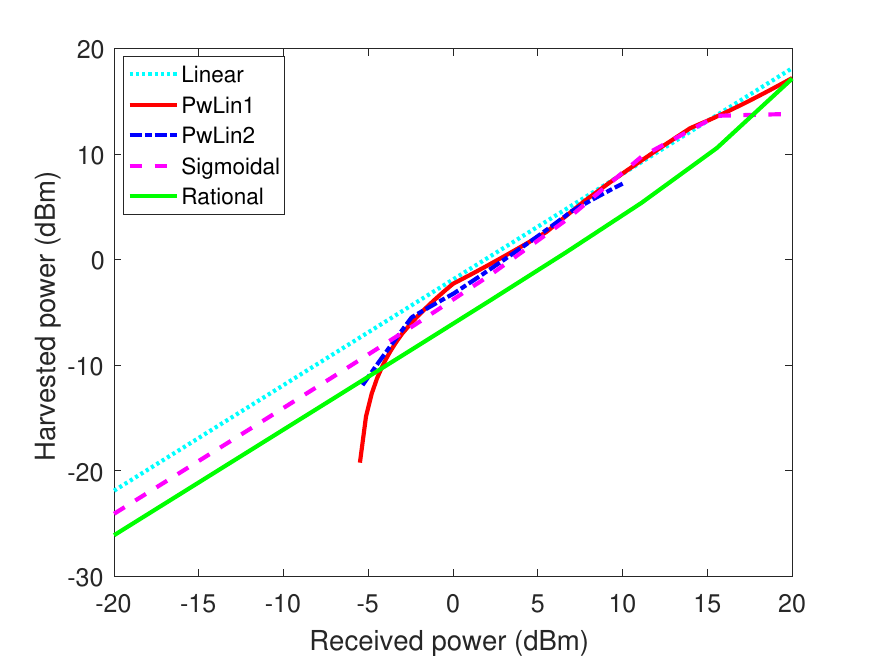}
	\caption{Comparison of various harvesting efficiency models. The first two piece-wise linear models, abbreviated as `PwLin', are reproduced from \cite{mishraICC2017, sarmaAccess2019}. The normalised sigmoidal and the rational models from \cite{boshkovskaTCOM2017, chenTVT2016}, respectively are plotted along with the linear ($\eta=0.7$) model.}
	\label{fig:harvVsRecvComp}
\end{figure}

The harvesting efficiency, denoted as $\eta(\centerdot)$, is a function of the received power, as established in the energy harvesting circuit literature \cite{valentaMM2014}. As a result, the harvested power, $P_h$ is related to the received power $P_r$ through the following equation
\begin{align}\label{eq:nonlineffcncy}
P_{h} = \eta(P_r)P_r.
\end{align}
The indirect feedback scheme proposed in this article requires the harvested power to be a monotonic function of the received power, i.e., $dP_h/dP_r > 0$. In other words, if $P_{r_1}$, $P_{r_2}$ are two received power values, and $P_{h_1}$, $P_{h_2}$ are corresponding harvesting power values, respectively, then $P_{h_1} < P_{h_2}$ if and only if, $P_{r_1} < P_{r_2}$. It is apparent that if $\eta(\centerdot)$ is a constant or a non-decreasing function of $P_r$, the above requirement is easily satisfied. Further, one can easily verify various model considered in the literature for estimating harvested power from the received power, e.g., piece-wise linear \cite{mishraICC2017, alevizosTWC2018,sarmaAccess2019}, normalised sigmoid function \cite{boshkovskaTCOM2017} and the rational function \cite{chenTVT2016}, meet the required criterion in their respective non-saturation region. A comparison of some of these models is depicted in Fig. \ref{fig:harvVsRecvComp}. %\txtred{Figure \ref{fig:harvVsRecvComp} compares some of these models.} % until the harvester enters the saturation region.} %in Strictly speaking, we demand the harvesting efficiency to satisfy the following relation: \footnotetext{first order derivative of a monotonically increasing function is greater than $0$}
%\begin{align}\label{eq:monotonicityCons}
%\frac{\eta(x)}{x} > \frac{d}{dx}\eta(x), \ \ \ \ 0 \le x \le P_{r,max},
%\end{align}
%where $P_{r,max}$ is the maximum received power allowed at the input of the harvester circuit. 
%The output of the rectifier circuit ends at an energy storage block where the electric charge gets accumulated. Though a practical energy storage block might be equipped with multi-stage charge pump circuits, e.g., Cockcroft-Walton or Dickson charge pumps (cf. \cite{valentaMM2014}), or a boost converter, for the ease of exposition we shall assume a simple R-C circuit model. 
\subsubsection*{Constant power source and energy storage}
A super-capacitor acting as energy storage accumulates electric charge from the signal processed by the rectifier-filter block. In practical harvesting systems, multi-stage charge pump circuits, e.g., Cockcroft-Walton or Dickson charge pumps \cite{valentaMM2014}, or boost converters precede the energy storage block for the improved charge delivery. %However, for the ease of exposition, we assume a simple R-C circuit model for the rectifier and energy storage blocks .  
However, similar to \cite{abeywickramaJSTSP2021}, we model the energy storage as a series R-C circuit for ease of exposition (refer to Fig. \ref{fig:systemOverview}(c)). Further, we assume that the output of the rectifier block acts as a constant power source, proposed and validated in \cite{mishraTCSIIB2015}, and drives the R-C circuit.  

\subsubsection*{Time-to-recharge}
\begin{figure}[!h]
	\centering
	\includegraphics[scale=0.45]{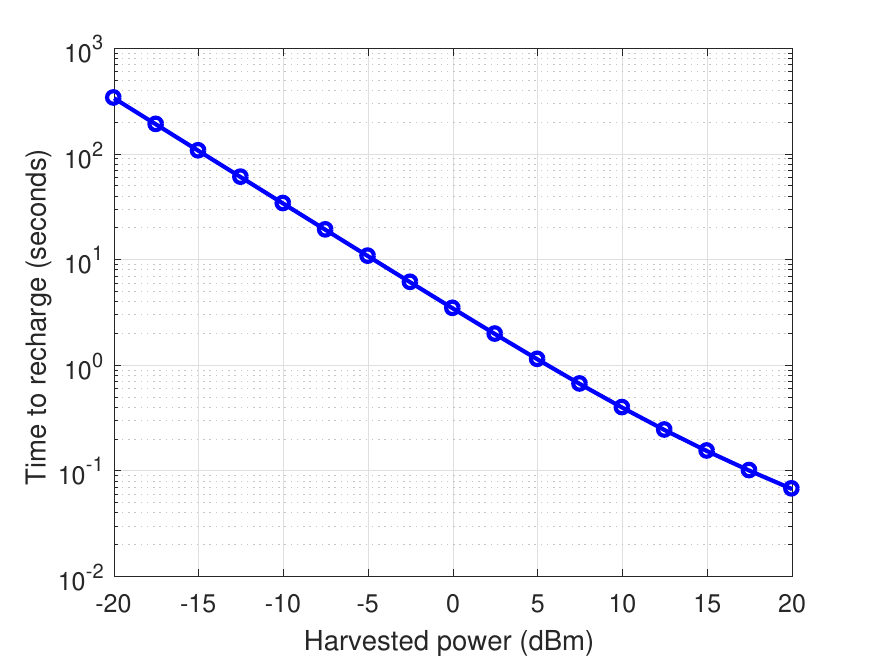}
	\caption{Time-to-recharge vs harvested power obtained from \eqref{eq:timeToRecharge} by considering $R = 100 \Omega$, $C = 1$mF, $q_0 = 1.5$ milli Coulombs, and $q_m = 3$ mC. }
	\label{fig:ttrVsHarvPwr}
\end{figure}

The constant power source R-C circuit based energy storage system was considered in \cite{sarmaAccess2019} to obtain a closed-form expression for the \emph{time-to-recharge} of an energy harvester. It was shown that the time-to-recharge of an energy harvester for a fixed received power (at the input of the R-C circuit) is
%Further, we consider a constant power source model, proposed and validated in \cite{mishraTCSIIB2015}, for the received RF signal at the energy harvester. 
%Further, \txtred{relying on the results} provided in \cite{mishraTCSIIB2015}, we shall model the\cmnt{rectified signal} rectifier and the preceding RF block including antenna as constant power source for the succeeding energy storage block. % \cite{mishraTCSIIB2015}.
%The constant power simplified R-C circuit, initially proposed and \txtred{cross-checked} against the experimental data obtained from powercast P2110 evaluation board in \cite{mishraTCSIIB2015}, was considered in \cite{sarmaAccess2019} to obtain a generalized closed-form expression for the \emph{time-to-recharge} an energy harvester before it is ready to transmit an information message. Assuming a fixed harvested power, it was shown  that the time-to-recharge the super-capacitor embedded on the energy harvester from an initial charge level, $q_0$, to a desired charge level, $q_m$, is given by
\begin{align}\label{eq:timeToRecharge}
\!t_{tr}(q_m,q_0,P_{h})\! =\! \frac{RC}{2}\! \ln\! \left(\frac{ Y(q_m, P_h)\exp(Y(q_m, P_h))}{ Y(q_0, P_h)\exp(Y(q_0, P_h))} \right),
\end{align}
where $Y(q,P) =  \frac{(q+\sqrt{q^2+4PRC^2})^2}{4PRC^2}$, $P_{h}$ is the input power to the R-C circuit, $R$ and $C$ denote the resistance and capacitance of the series R-C (energy storage) circuit, and $q_0, q_m$ are the initial and desired charge levels, respectively. Intuitively, a larger value of $P_h$ will result in a smaller $t_{tr}$ and the same can be seen \cmnt{shown from \eqref{eq:timeToRecharge} and}from Fig. \ref{fig:ttrVsHarvPwr} as well. % it is evident that a higher value of $P_h$ will result in a smaller $t_{tr}$. 

%. \txtred{By referring to the basic principles of circuit theory,} one can easily establish that a higher value of $P_h$ will result in a smaller $t_{tr}$. Additionally, we assume a nonlinear harvesting efficiency function that maps the received power, $P_r$, to the harvested power through the following relation %Now coming to the harvesting efficiency,

\subsubsection*{Indirect feedback from the time-to-recharge}
From \eqref{eq:rxPower} and \eqref{eq:nonlineffcncy}, we identify the relation between a beamforming vector and the resultant harvested power. By expressing the harvested power as a function of the beamforming vector, for given $q_0, q_m$, we obtain the following functional relation between the time-to-recharge and the corresponding beamforming vector from \eqref{eq:timeToRecharge}
%Kindly note that in the above expression the harvesting efficiency, $\eta(\centerdot)$, is a function of the received power, which has already been established in the energy harvesting circuit related literature. Now, from \eqref{eq:rxPower}, \eqref{eq:timeToRecharge} and \eqref{eq:nonlineffcncy} we can \txtred{draw} a relation between the chosen beamforming vector, $\bw$, and time-to-recharge the energy harvester, $t_{tr}$, which is provided below in a \txtred{functional form} for the ease of exposition.
\begin{align}\label{eq:ttrBmfrm}
t_{tr-bm} (\bw) = t_{tr}(q_m, q_0, P_h(\bw)) = f_{ttr}(|\bh^\dagger\bw|). 
\end{align}  
The indirect feedback about a beamforming vector is obtained through \eqref{eq:ttrBmfrm}---the closer a vector is to the instantaneous channel gain in cosine similarity, the smaller the time-to-recharge. Additionally, if \eqref{eq:nonlineffcncy} is satisfied, we can find the harvested power $P_h$ and subsequently, $|\bh^\dagger\bw|$ through an inverse mapping 
$f_{ttr}^{-1}: \R_+ \to \R_+$, such that 
\begin{align}\label{eq:absDotProdVal}
f_{ttr}^{-1}(t_{tr-bm}) = |\bh^\dagger\bw|.
\end{align} 

In the following subsections, we describe how an energy transmitter can obtain indirect feedback about the beamforming vector employed and subsequently improve the power delivery by adopting our proposed algorithm.%we describe the transmission \txtred{scheme} an energy transmitter must follow to utilize the indirect feedback for improving the power delivery at the energy receiver. 
%In the blind channel estimation process, described in the following subsection, we shall use this inverse mapping to estimate the channel gain from the $t_{tr}$, which is obtained by overhearing the transmission of the energy harvester device.
%
%\cmnt{\subsection*{Antenna weight selection and channel estimation} }
\subsection{Transmission phases and their roles}
To capitalize on the indirect feedback received from the energy receiver, we propose dividing the total stretch of energy transmission into two phases: 1) \emph{feedback acquisition and processing} and 2) \emph{energy beamforming transmission}, occurring alternately. Fig. \ref{fig:systemOverview}(d) illustrates the sequencing of these two phases in an ongoing energy transmission-harvesting process. The activities carried out by the energy transmitter and receiver in these phases are briefly described in the following subsections. 
%\txtred{To elucidate} the blind channel estimation process and the subsequent energy beamforming by the energy transmitter, we divide the total stretch of the energy transmission in alternate occurrences of two phases, namely, \emph{estimation-focused energy transmission} and \emph{energy beamforming transmission}. The fig. \ref{fig:timeSlotBlckDgrm}(a) illustrates the sequencing of these two phases in an ongoing energy transmission-harvesting process. \txtgry{Below we briefly describe these two phases.}

%\subsubsection*{Estimation-focused energy transmission phase} 
\subsubsection*{Feedback acquisition and processing (FAP)}
In this phase, as the name suggests, the energy transmitter acquires (indirect) feedback regarding a set of known beamforming vectors \cmnt{antenna weight vectors }by overhearing the (information) transmissions of the energy harvester. Essentially, the energy transmitter measures the time-to-recharge for these vectors one by one and feeds those measurements to Algorithm \ref{algo:blindChnlEst}, described in the following section. The measurement can be done by starting a timer whenever the transmitter employs a new \cmnt{antenna weight }vector and stopping it upon overhearing the energy harvester's transmission. We are assuming that the measurement taken in this manner is error-free. For ease of illustration, we divide this phase into several slots, each corresponding to a beamforming \cmnt{antenna weight }vector. Needless to say, the duration of these slots will differ as different \cmnt{antenna weight } beamforming vectors are employed in every slot. 

%In this phase, as the name suggests, the energy transmitter estimates the channel gain vector by measuring the time-to-recharge for a set of known antenna weight vectors and feeding those measurements to the blind channel estimation algorithm. % and finding the time-to-recharge for that vector by overhearing the transmission of the energy harvester, and repeating this process for a finite number of iterations as prescribed by the blind estimation algorithm. %before transmitting the energy signal % after setting its antenna weights according to a beamforming vector, which is either selected from a set of orthonormal vectors or obtained by combining a subset of those vectors. 
%For the ease of illustration, we divide this phase in several slots, where in each slot the energy transmitter employs a different beamforming vector. As shown in the fig. , the transmitter calculates the duration of a slot by starting a timer when it begins transmission by choosing a particular beamforming vector and stops that timer when it overhears the transmission of the energy harvester device. The time duration calculated is then fed to the blind estimation algorithm, which is presented in the following section. 
%It is apparent that the duration of these slots will be different from each other as we are changing the beamforming vectors each time. 

\subsubsection*{Energy beamforming transmission (EBT)} 
This phase ensues the FAP phase once the optimal beamforming vector is calculated by the Algorithm \ref{algo:blindChnlEst}. The optimal beamforming vector is employed to deliver power to the energy harvester in this phase. 
%starts once the time-to-recharge for all known antenna weight vectors are acquired and the \txtred{channel estimation algorithm} presented in the following section produces a suitable beamforming vector. This beamforming vector is employed by the energy transmitter to deliver power to the harvesting device in this phase. 
As the beamforming vector remains unchanged throughout this phase, slotting is not required. However, the energy transmitter continues measuring the time-to-recharge by finding the interval between two consequent information transmissions. 

%This phase starts once the channels estimation process is over. The energy transmitter, in this phase, performs energy beamforming by setting its antenna weights according to the estimated channel gain vector. As the beamforming vector remains unchanged throughout this phase, there is no need for dividing this phase in slots as it was done for the estimation-focused transmission phase. However, in this phase also, the energy transmitter continues measuring the time-to-recharge by finding the interval between two consequent information message transmissions. 
Note that in the proposed indirect feedback scheme, the energy harvester harvests energy from the transmission of the energy transmitter in both phases. 

\section{Finding the optimal beamforming vector from the indirect feedback}\label{sec:algo}
In this section, we present an algorithm that finds the optimal beamforming vector from the feedback collected in the FAP phase. In a nutshell, the proposed algorithm finds the normalized and equally rotated coefficients of the channel gain in an orthonormal basis of $\C^N$ from the indirect feedback of the linear combination of vectors from the chosen basis. 
	\begin{algorithm}
	\caption{ Algorithm for finding the optimal beamforming vector}
	\label{algo:blindChnlEst}
	\begin{algorithmic}[1]
		\REQUIRE $\mb{Q}, \phi_1$, $\phi_2$ % $num\_antennas$, number of antennas on the transmitter \\ %$\tau_i, i \in \{1,2,\ldots M\}$
%	 (angles for finding the relative angles)
		\ENSURE $\bw_{opt}$ % $( = \bh e^{j\theta}/||\bh||)$ 
%		\STATE Generate a unitary (circulant) matrix, $\mb{Q}$ having $num\_antennas$ columns.  
		\STATE Initialization: $\phi_1 \gets \pi/4$, $\phi_2 \gets 7\pi/4$.
		\FOR{$i=1$ \TO $N$} 
		% // number of transmission required $= N$
%		\STATE $\mb{w} \leftarrow \mb{Q}(:,i)$    \txtgry{// using MATLAB notation for columns of a matrix }
		\STATE Set $\mb{Q}(:,i)$ as beamforming vector and find $\tau_i$ from \eqref{eq:absDotProdVal} %\txtgry{// to be used in one transmission slot} %from the time taken to charge the capacitor. 
		\ENDFOR
		\STATE  $\bw_{opt} \leftarrow \mb{Q}(:,1)$
		\STATE  $dot\_prod\_\bw_{opt} \leftarrow \tau_1$
		\FOR{$i=1$ \TO $N - 1$}
		\STATE $\alpha_1 \leftarrow dot\_prod\_\bw_{opt}$, $\alpha_2 \leftarrow \tau_{i+1}$
		\STATE $\mu \leftarrow \alpha_1^2 + \alpha_2^2$
		\STATE $\kappa_1 \leftarrow (\alpha_1^4 + \alpha_2^4)/\mu$, $\kappa_2 \leftarrow \alpha_1\alpha_2/\mu$
		\STATE $\mb{w}_1 \leftarrow (\alpha_1\bw_{opt} + e^{j\phi_1}\alpha_2\mb{Q}(:,i+1))/\sqrt{\mu}$ %            \txtgry{// $j^2 = -1$}
		\STATE Set $\mb{w}_1$ as beamforming vector and find $\tilde{\tau}_1$  %\txtgry{// one more transmission slot} 
		\STATE $\mb{w}_2 \leftarrow (\alpha_1\bw_{opt} + e^{j\phi_2}\alpha_2\mb{Q}(:,i+1))/\sqrt{\mu}$
		\STATE Set $\mb{w}_2$ as beamforming vector and find $\tilde{\tau}_2$   % \txtgry{// one more transmission slot}
		%\STATE $\beta_1 \leftarrow \frac{\tilde{\tau}_1^2 -\kappa_1}{\sqrt{2}\kappa_2}$
		%\STATE $\beta_2 \leftarrow (\tilde{\tau}_2^2 -\kappa_1)/(\sqrt{2}\kappa_2)$
		\STATE $\gamma_R \leftarrow \frac{\tilde{\tau}_1^2 + \tilde{\tau}_2^2 - 2\kappa_1}{2\sqrt{2}\kappa_2}$, $\gamma_I \leftarrow \frac{\tilde{\tau}_1^2 - \tilde{\tau}_2^2}{2\sqrt{2}\kappa_2}$
		\STATE $\theta_1 \leftarrow \arctan\left(\frac{-\gamma_I}{\gamma_R}\right)$, $\theta_2 \leftarrow \arctan\left(\frac{\gamma_I}{-\gamma_R}\right)$
		\IF{$(\gamma_R\cos\theta_1- \gamma_I\sin\theta_1) > 0$}
		\STATE $\bw_{opt} \gets (\alpha_1\bw_{opt}\! +\! e^{j\theta_1}\alpha_2\mb{Q}(:,i\!+\!1))$, \\ $dot\_prod\_\bw_{opt} \!\!\gets  \!\!\sqrt{\kappa_1 \!\!+ \!\!2\kappa_2(\gamma_R\cos\theta_1 \!\! - \!\! \gamma_I\sin\theta_1)}$ % $\theta_{i+1}^* \leftarrow \theta_1$,
		\ELSE
		\STATE $\bw_{opt} \gets (\alpha_1\bw_{opt} + e^{j\theta_2}\alpha_2\mb{Q}(:,i+1))$, \\ $dot\_prod\_\bw_{opt} \!\! \gets  \!\!\sqrt{\kappa_1 \!\!+ \!\!2\kappa_2(\gamma_R\cos\theta_2  \!\!-  \!\!\gamma_I\sin\theta_2)}$ %$\theta_{i+1}^* \leftarrow \theta_2$,
		\ENDIF 
		\ENDFOR  
		\RETURN $\bw_{opt}$
	\end{algorithmic}	
	\end{algorithm}

Assuming that $\{\bq_1, \bq_2,\ldots, \bq_N \}$ is an ordered orthonormal basis of $\C^N$, the unknown channel vector $\bh$ can be expressed as %  The channel gain vector $\bh$ belongs to $\C^N$, it must be expressible as a linear combination of the vectors of a basis of $\C^N$. The column vectors of matrix $\mb{Q}$ form one such orthonormal basis. For notational simplicity, let us denote $\mb{Q}(:,i)$ by $\bu_i$ Hence, 
\begin{align}\label{eq:chnlGainLinComb}
\bh = \sum\limits_{i=1}^{N}\zeta_i\bq_i, \text{ where } \zeta_i \in \C, \text { for } i \in \{1,2,\ldots, N\}.
\end{align}  

%Similar to the conventional pilot-based channel estimation schemes  
To ease exposition, we collect these basis vectors in a matrix, $\mb{Q}$ and refer to them using the column index notation, $\mb{Q}(:,i)$. Recalling that $\bq_i$ is orthogonal to $\bq_j$, $j \ne i$, one can easily verify that $\mb{Q}^\dagger\bh = \boldsymbol{\zeta}$, where $\boldsymbol{\zeta} = [\zeta_1, \zeta_2,\ldots,\zeta_N]^T$. %comprises $\zeta_i$.}

When $\bq_i$ is the beamforming vector, the absolute value of the dot product of the channel gain and the beamforming vector (denoted by $\tau_i$ in the Algorithm \ref{algo:blindChnlEst}) is  %and transmit the energy signal, we obtain $\tau_i$ using \eqref{eq:absDotProdVal}. The $\tau_i$ are absolute value $\zeta_i$, as can be seen below 
\begin{align}
|\bh^\dagger\bq_i| = \big|\big(\sum_{k=1}^{N}\zeta_k\bq_k\big)^\dagger\bq_i\big| = |\zeta_i|. \label{eq:absDotProd}%, \ \ \ [\because \bu_k^\dagger\bu_i = 1, \text{ when } k=i; 0, \text{ otherwise }].%[\because \mb{Q}^\dagger\mb{Q} = \mb{I}_{N}]
\end{align}

Note that in several conventional channel estimation schemes, $\zeta_i$ ({its noisy version to be precise}) is available at the receiver as the received signal is a linear function of $\bh^\dagger\bw$. Unlike those schemes, the proposed algorithm tries to estimate the channel from the absolute values of $\zeta_i$, and therefore, requires additional feedback as well as computation. Consequently, the estimated channel gain is a rotated and scaled version of the actual one, which is sufficient for finding the optimal beamforming vector as pointed out in \eqref{eq:rxPower}. The optimal beamforming vector produced by the algorithm is 
%If we have to find the exact $\bh$, then along with $|\zeta_i|$, we also need to find $\angle\zeta_i$, as $\zeta_i$ can be expressed in the polar form as $ |\zeta_i|e^{j\angle\zeta_i}$. However, as mentioned earlier, any rotated and scaled version of the channel gain vector will suffice for finding the optimal beamforming vector. Hence, we are considering the following vector as an estimate for the channel gain vector
\begin{align}\label{eq:chnlEst}
%\bh_{est} = \sum\limits_{i=1}^{N}|\zeta_i|\bq_ie^{j(\angle\zeta_i -\angle\zeta_1)}/\sqrt{\sum_{i=1}^{N}|\zeta_i|^2}.
\bw_{opt} = \sum\limits_{i=1}^{N}\frac{|\zeta_i|}{\sqrt{\sum_{i=1}^{N}|\zeta_i|^2}}\bq_ie^{j(\angle\zeta_i -\angle\zeta_1)}. %, \text{ where } \zeta_i = |\zeta_i|e^{j\angle\zeta_i}. %\frac{1}{\sqrt{\sum\limits_{i=1}^{N}|\zeta_i|^2}}
\end{align}
%\txtblu{where $\nu_i = |\zeta_i|/\sqrt{\sum_{i=1}^{N}|\zeta_i|^2}$. 
From the above expression, it is apparent that the phase difference between the estimate and the actual channel vector is $\angle\zeta_1$.  

The proposed algorithm has two for loops---the first one for finding the absolute dot products (cf. \eqref{eq:absDotProd}), and the second for calculating relative angles (cf. \eqref{eq:chnlEst}) corresponding to each $\zeta_i, i=2,\ldots, N$. Once $\tau_i (= |\zeta_i|)$ are obtained by employing $\bq_i$ as beamforming vectors in steps 3 - 6, the Algorithm \ref{algo:blindChnlEst} finds the angle between $i$th coefficient and the first one, i.e., $\angle\zeta_i -\angle\zeta_1$ in an iterative manner. To illustrate this segment (steps 7 - 23) of the algorithm, let us consider the following unit-norm beamforming vector formed by the linear combination of two known unit-norm vectors, $\mb{v}_1, \mb{v}_2$:
  \begin{equation}\label{eq:bfvecTheta}
 \tilde{\bw}_\theta = \frac{|\bh^\dagger\mb{v}_1|\mb{v}_1 + e^{j\theta}|\bh^\dagger\mb{v}_2|\mb{v}_2}{\sqrt{|\bh^\dagger\mb{v}_1|^2 + |\bh^\dagger\mb{v}_2|^2}}. %(\mb{v}_1, \mb{v}_2, \theta )
 \end{equation} %\tilde{\bw}_\theta = (|\zeta_1|\bq_1 + |\zeta_2|e^{j\theta}\bq_2)/(|\zeta_1|^2 + |\zeta_2|^2)$.
% where $||\mb{v}_1|| = ||\mb{v}_2|| = 1$.
% by considering first two vectors of the chosen orthonormal basis.   }
%The angle $\angle\zeta_i-\angle\zeta_1$ corresponds to the relative angle of the $i$th vector with respect to the first vector of the chosen basis. \cmnt{ in the channel vector estimate under consideration} To illustrate the relative angle calculation, let us consider the first two vectors of the chosen basis. Also, 
In the first iteration, we set $\mb{v}_1 = \bq_1$, $\mb{v}_2 = \bq_2$ and generate intermediate unit-norm beamforming vectors of the following form \cmnt{ $\tilde{\bw}_\theta$ is formed by adding $\bq_2e^{j\theta}$ to $\bq_1$ and then normalizing it. Hence,} $\tilde{\bw}_\theta = (|\zeta_1|\bq_1 + |\zeta_2|e^{j\theta}\bq_2)/(|\zeta_1|^2 + |\zeta_2|^2)$. %Now, if this vector is used for transmitting the energy signal, then 
The numerator of absolute dot product for this beamforming vector can be evaluated as
\begin{align}
& \bigg|\big(\sum\limits_{i=1}^{N}\zeta_i\bq_i\big)^\dagger(|\zeta_1|\bq_1 + e^{j\theta}|\zeta_2|\bq_2)\bigg|^2 \nonumber \\
& = \big|\zeta_1^*|\zeta_1| + \zeta_2^*|\zeta_2|e^{j\theta}\big|^2 \nonumber \\
%& = \left((\zeta_1^*|\zeta_1| + \zeta_2^*|\zeta_2|e^{j\theta})^*(\zeta_1^*|\zeta_1| + \zeta_2^*|\zeta_2|e^{j\theta})\right) \\
& = |\zeta_1|^4 + |\zeta_2|^4 + |\zeta_1||\zeta_2|(\zeta_1\zeta_2^*e^{j\theta} + \zeta_1^*\zeta_2e^{-j\theta}) \label{eq:dotProdTheta}
\end{align}
%\begin{align}
%%|\bh^\dagger\bw_\theta^{(1,2)}|^2 
%& |\bh^\dagger\tilde{\bw_\theta}|^2 \nonumber\\
%& = \frac{1}{(|\zeta_1|^2 + |\zeta_2|^2)}\left|\left(\sum\limits_{i=1}^{N}\zeta_i\bq_i\right)^\dagger(|\zeta_1|\bu_1 + e^{j\theta}|\zeta_2|\bu_2)\right|^2 \nonumber\\
%& =  \frac{1}{|\zeta_1|^2 + |\zeta_2|^2}|\zeta_1^*|\zeta_1| + \zeta_2^*|\zeta_2|e^{j\theta}|^2 \nonumber \\
%& =  \frac{1}{|\zeta_1|^2 + |\zeta_2|^2}\left((\zeta_1^*|\zeta_1| + \zeta_2^*|\zeta_2|e^{j\theta})^*(\zeta_1^*|\zeta_1| + \zeta_2^*|\zeta_2|e^{j\theta})\right) \nonumber \\
%& =  \frac{1}{|\zeta_1|^2 + |\zeta_2|^2}\left(|\zeta_1|^4 + |\zeta_2|^4 + |\zeta_1||\zeta_2|(\zeta_1\zeta_2^*e^{j\theta} + \zeta_1^*\zeta_2e^{-j\theta})\right) \label{eq:dotProdTheta}
%%& = \frac{|\zeta_1|^4 + |\zeta_2|^4}{|\zeta_1|^2 + |\zeta_2|^2} + \frac{|\zeta_1||\zeta_2|}{|\zeta_1|^2 + |\zeta_2|^2}2\Re\{\zeta_1\zeta_2^*e^{j\theta}\}\\
%\end{align}
To find the $\theta$ that maximizes (or minimizes) the absolute dot product, we differentiate \eqref{eq:dotProdTheta} with respect to $\theta$ and equate to $0$, i.e., $\zeta_1\zeta_2^*e^{j\theta} - \zeta_1^*\zeta_2e^{-j\theta} = 0$. In other words, $\theta$ should be such that the imaginary part of $\zeta_1\zeta_2^*e^{j\theta}$ is zero, i.e.,  $\Im\{\zeta_1\zeta_2^*e^{j\theta}\} = 0$. After simple algebraic manipulation, we find the following expression to calculate the desired $\theta$ % \cmnt{, let us denote it by $\wtht$, } must satisfy the following relation 
\begin{align}\label{eq:optTheta}
\tan \theta = - \frac{\Im\{\zeta_1\zeta_2^*\}}{\Re\{\zeta_1\zeta_2^*\}}.
\end{align}
For notational simplicity, we denote $\Im\{\zeta_1\zeta_2^*\}, \ \Re\{\zeta_1\zeta_2^*\}$ by $\gamma_{I}, \gamma_{R}$, respectively in the subsequent paragraphs as well as in the Algorithm \ref{algo:blindChnlEst}. 
As the negative sign in the above equation can be associated with either the numerator or the denominator, there are two possible values for $\theta$. Let us denote them by $\theta_1$ and $\theta_2$. 
%\cmnt{ potential candidates for $\widehat{\theta}$. we obtain two $\wtht$ values, namely, $\wtht_1$ and $\wtht_2$, upon solving it. 
Note that one of them is the maximizer of $|\bh^\dagger\twth|^2$ , whereas, the other one is the minimizer. By putting these two values in the second derivative of $|\bh^\dagger\twth|^2$ with respect $\theta$ we can identify the maximizer---the second derivative evaluated at the maximizer will return a negative value (step 17 of Algorithm \ref{algo:blindChnlEst}). 

As we need \cmnt{$\Im\{\zeta_1\zeta_2^*\}, \ \Re\{\zeta_1\zeta_2^*\}$, denoted by} $\gamma_{I}, \gamma_{R}$ \cmnt{, respectively in the algo \ref{algo:blindChnlEst},} to find $\theta_1, \theta_2$ from \eqref{eq:optTheta}, the algorithm obtains the absolute dot products for two specific intermediate unit-norm beamforming vectors, formed using \eqref{eq:bfvecTheta} and by choosing $\theta=\pi/4, 7\pi/4$ (steps 11 - 14 of Algorithm \ref{algo:blindChnlEst}). The corresponding absolute dot product values, \cmnt{ and employ them in energy transmission in the consequent slots to find the corresponding absolute dot product values, }denoted by $\tilde{\tau}_1$ and $\tilde{\tau}_2$, respectively in the Algorithm \ref{algo:blindChnlEst}, can be expressed as%. By replacing $\theta$ by $\pi/4$ \cmnt{ and $7\pi/4$} in \eqref{eq:dotProdTheta}, we obtain the following expressions
\begin{subequations} \label{eq:dotProdIBf}
\begin{align}
\tilde{\tau}_1^2  &= |\bh^\dagger\tilde{\bw}_{\frac{\pi}{4}}|^2 = \kappa_1 + 2\kappa_2\Re\{\zeta_1\zeta_2^*e^{j\pi/4}\}, \\ % \label{eq:dotProdIBf1} \\
\tilde{\tau}_2^2  &= |\bh^\dagger\tilde{\bw}_{\frac{7\pi}{4}}|^2 = \kappa_1 + 2\kappa_2\Re\{\zeta_1\zeta_2^*e^{j7\pi/4}\}, %\label{eq:dotProdIBf2}
\end{align}
\end{subequations}

where $\kappa_1\!\! = \!\! (|\zeta_1|^4\!\! +\!\! |\zeta_2|^4 )/(|\zeta_1|^2\!\! +\!\! |\zeta_2|^2)$ and $\kappa_2 = |\zeta_1||\zeta_2|/(|\zeta_1|^2 \!+\! |\zeta_2|^2)$. 
On expanding and rearranging \eqref{eq:dotProdIBf}, we obtain the following linear equations in  $\gamma_{R}, \gamma_{I}$ % $\Re\{\zeta_1\zeta_2^*\}$ and  $\Im\{\zeta_1\zeta_2^*\}$
\begin{subequations}\label{eq:linEqnZeta}
\begin{align}
%\Re\{\zeta_1\zeta_2^*\} - \Im\{\zeta_1\zeta_2^*\} & 
\gamma_{R} - \gamma_{I} & = (\tilde{\tau}_1^2 - k_1)/(\sqrt{2}\kappa_2), \\ % \label{eq:linEqnZetas1} \\
%\end{align}
%Similarly, by replacing $\theta$ by $7\pi/4$ in \eqref{eq:dotProdTheta}, we obtain 
%\begin{align}
%\Re\{\zeta_1\zeta_2^*\} + \Im\{\zeta_1\zeta_2^*\} 
\gamma_R + \gamma_I & = (\tilde{\tau}_2^2 - k_1)/(\sqrt{2}\kappa_2). %\label{eq:linEqnZetas2}
\end{align}
\end{subequations}
Upon solving \eqref{eq:linEqnZeta}, \cmnt{and \eqref{eq:linEqnZetas2} together} $\gamma_R, \gamma_I$ are obtained and subsequently $\theta_1, \theta_2$ are calculated \cmnt{$\Im\{\zeta_1\zeta_2^*\}, \Re\{\zeta_1\zeta_2^*\}$ and  that are used later to calculate $\widehat{\theta}$ from \eqref{eq:optTheta}} (steps 15 - 16). 

The $\theta$ corresponding to the maximizer is identified with the help of the second derivative of the numerator of \eqref{eq:bfvecTheta}. The optimal beamforming vector, denoted by $\bw_{opt}$, is calculated by evaluating \eqref{eq:bfvecTheta} at that $\theta$.  %using $\widehat{\theta}$ in the expression of $\tilde{\bw}_\theta$. 
In the next iteration, the algorithm calculates the relative angle between current $\bw_{opt}$ and $\bq_3$ and subsequently updates the $\bw_{opt}$ by repeating the procedure mentioned above. Note that the absolute value of the dot product of $\bh$ and current $\bw_{opt}$ can be obtained from \eqref{eq:dotProdTheta} and is expressed as (steps 18, 20)
\begin{equation*}\label{eq:optBfAbsDotProd}
|\bh^\dagger\bw_{opt}| = \sqrt{\kappa_1 +2\kappa_2(\gamma_R\cos\theta_2  - \gamma_I\sin\theta_2)}.
\end{equation*}

%We updated estimate of the channel gain vector after this step is $\bw_{\widehat{\theta}}^{(1,2)}$. The algorithm, in the next iteration, repeats the same process with $\bw_{\widehat{\theta}}^{(1,2)}$ as the first vector, and $\bu_3$ as the second one. 
\subsubsection*{Number of iterations} 
The algorithm returns the optimal beamforming vector in $3N -2$ iterations, where in the initial $N$ iterations the absolute dot products of the channel gain and the basis vectors are obtained. The following $2(N-1)$ iterations are spent on estimating the relative angles (maximizer) between the current optimal beamforming vector and the next basis vector in the considered ordered basis.   
% As each iteration of this step spans over two consecutive slots, \cmnt{this step will run for} $2(N-1)$ slots of different lengths are needed to complete this block. 
 %Once the algorithm has found an estimate for the channel gain vector in $3N-2 \ \ (= N + 2N - 2)$ slots, the energy beamforming transmission phase will begin. 
 
\subsection*{Limitation of the proposed algorithm and a pragmatic solution}

For various real-world time-sensitive applications, timely information updates from wireless sensor nodes are critical to the usefulness of a deployed system. Therefore, a shorter feedback acquisition and processing phase is preferred. However, as the channel gain takes arbitrary values, it is indeed possible that the absolute dot products of the channel gain and one or more basis vectors are negligibly small. For example, when the channel gain lies in the null space of the matrix formed by a set of basis vectors. In such cases, the harvested power will be essentially zero, resulting in no information transmission (indirect feedback) from the energy harvester.  
	
From Fig. \ref{fig:ttrVsHarvPwr}, it is apparent that when the harvested power falls below -15 dBm, the time-to-recharge increases to 100 secs. To prevent undesirable delays in the indirect feedback and curtail the feedback acquisition and processing phase, we propose setting a time-limit for each slot of this phase. Whenever the indirect feedback is not received within a pre-decided time-limit, the transmitter switches to one of the previously used beamforming vectors, preferably, the one with minimum time-to-recharge. Once the feedback is obtained after switching to a different beamforming vector, the time-to-recharge is recorded and the algorithm picks the next vector from the basis.

We argue that it is indeed possible to find the optimal beamforming vector even when a time-limit is imposed on slots, albeit, at the cost of additional computation. Let us assume that the time-limit is exceeded when the transmitter picked $\bq_k$ as the beamforming vector. The transmitter then switched to a previously used basis vector $\bq_i$ and continued with the transmission. By denoting the harvested energy due to $\bq_i, \bq_k$ as $P_{h_i}, P_{h_k}$, respectively, we can write the following equations from  \eqref{eq:timeToRecharge}
\begin{subequations}\label{eq:ttr_timLim}
\begin{align}
t_{lim} & = t_{tr}(q_0,q_s,P_{h_k}), \\
t_{res} & = t_{tr}(q_s,q_m,P_{h_i}),
\end{align}
\end{subequations}
where $t_{lim}, t_{res}$ represent the duration of the time-limit and the additional time needed to obtain the feedback after switching to $\bq_i$, respectively and $q_s$ denote the intermediate charge level of the super capacitor. Recall that $\bq_i$ was used previously and therefore, time-to-recharge for $\bq_i$ is already known. $P_{h_i}$ can be calculated from that and subsequently $q_s$ can be evaluated from \eqref{eq:timeToRecharge}. Once $q_s$ is obtained, $|\zeta_k|$ is calculated from \eqref{eq:absDotProdVal}. Note that if the time-limit is exceeded for the first vector from the ordered basis, the algorithm switches to the second vector to obtain $t_{res}$ and then continues with the same vector to find $P_{h_2}$ and $|\zeta_2|$.

\subsubsection*{Remarks} If it is observed that $t_{res} = t_{tr}(q_s,q_m,P_{h_i}) \approx t_{tr}(q_0,q_m,P_{h_i})$, one can conclude that $P_{h_k} \approx 0$ and therefore, $\bq_k$ can be excluded while calculating the (approx) optimal beamforming vector from the basis vectors. Furthermore, reordering the basis vectors in the descending order of the $\tau$ values will ensure that time-limit is not exceeded while finding the relative angles. 

The time-limit can be incorporated in the Algorithm \ref{algo:blindChnlEst} by replacing the step 3 with ones mentioned in Algorithm \ref{algo:optBftimLim}. In the Algorithm \ref{algo:optBftimLim}, for the sake of brevity, we denote the set of vectors for whom the indirect feedback is obtained within the time-limit by $\mathcal{Q}_{btl}$. 
 
\begin{algorithm}
\caption{Modification in the optimal beamforming vector to incorporate a time-limit}
\label{algo:optBftimLim}
\begin{algorithmic}[1] 
\STATE Transmit using $\mb{Q}(:,i)$ as beamforming vector
\IF{time-limit is exceeded}
\IF{$\mathcal{Q}_{btl}$ is not empty}
\STATE Pick the vector with largest $\tau$ and continue transmission till the indirect feedback is received.
\STATE From \eqref{eq:ttr_timLim}, find the $\tau$ for the vector that resulted in exceeding the time-limit.
\ELSE
\STATE Select the next vector from the ordered basis and continue transmission till an indirect feedback is received or the time-limit is exceeded.
\ENDIF
\ENDIF   
%\IF{All the vectors exceeded the time-limit}
%\STATE Time-limit should be increased to obtain the optimal beamforming vector.
%\ELSE
%\STATE Find $\tau$ for remaining vectors using \eqref{eq:ttr_timLim}. %, if not done yet.
%\ENDIF 
\end{algorithmic}
\end{algorithm}

\section{Proposed Hardware}\label{sec:hrdwr}

\begin{figure}
	\centering
	\includegraphics[scale=0.500]{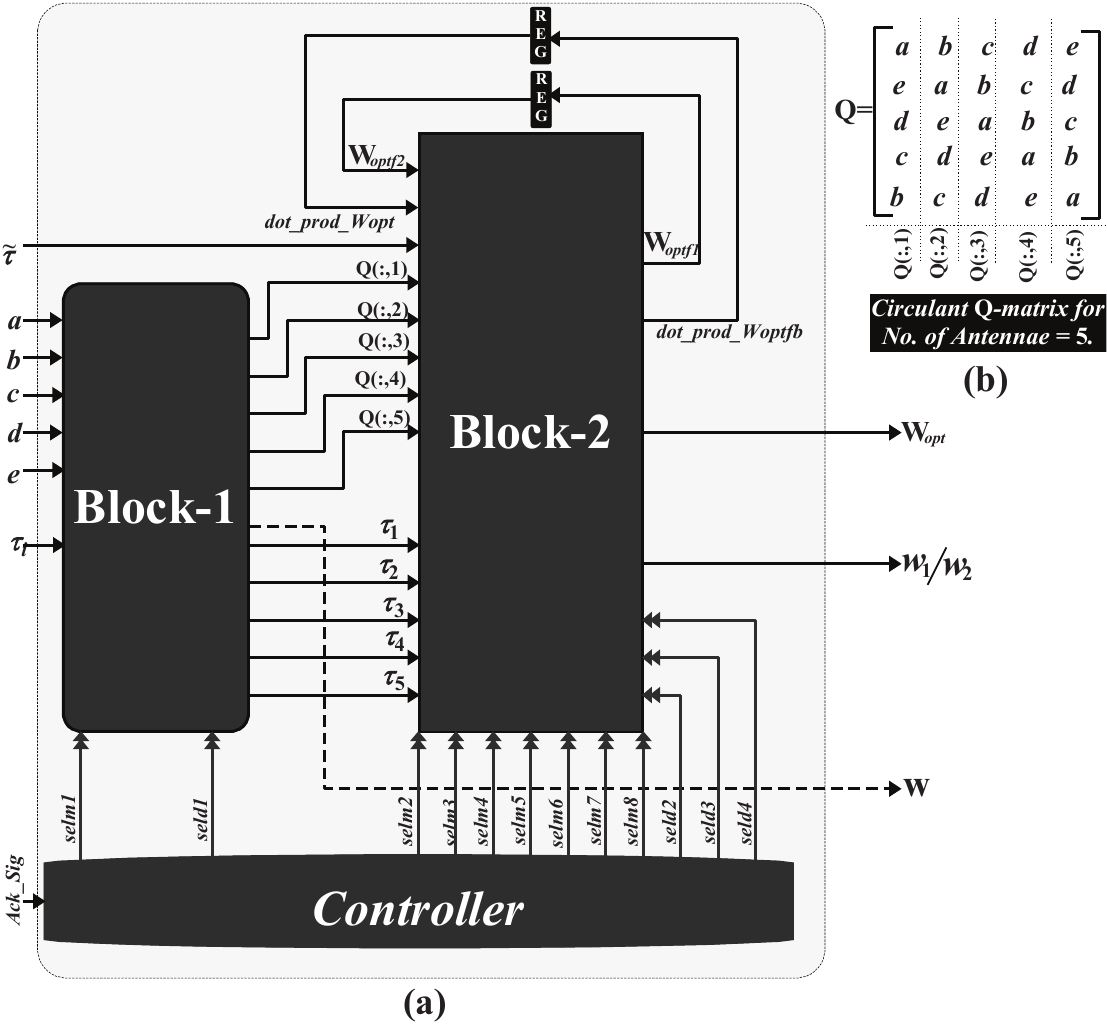}
	\caption{(a) Proposed system-level hardware-architecture of OEB for $N{=}5$ number of antennae (b) $\mathbf{Q}$ $\in$ $\mathbb{C}^{5{\times}5}$ circulant matrix containing 16-bit complex elements.}
	\label{est}\
\end{figure}

This section presents a new hardware architecture for the proposed Algorithm \ref{algo:blindChnlEst}. An overall design of such optimal energy beamformer (OEB) is shown in Fig. \ref{est} (a) that comprises three major parts: Block-1, Block-2, and controller. It has been design to support $N{=}5$ number of antennae with $\mathbf{Q}$ $\in$ $\mathbb{C}^{5{\times}5}$ circulant matrix, as shown in Fig. \ref{est} (b). In Fig. \ref{est} (a), Block-1 is designed to execute steps 2$-$4 from Algorithm \ref{algo:blindChnlEst} to generate the column-wise elements of $\mathbf{Q}$ matrix (i.e. $\mathbf{Q}$(:, \emph{i}) $\forall$ \emph{i} = \{1, 2, 3, 4, 5\}). These column vectors are iteratively transmitted as beamforming \textbf{w} vector in consecutive transmission slots for computing $\tau_{i}$ $\forall$ \emph{i} = \{1, 2, 3, 4, 5\} which are received and further transferred by Block-1 to Block-2 in the suggested architecture, as shown in Fig. \ref{est} (a).
Subsequently, Block-2 of the proposed OEB has been designed to perform the operations of steps 8$-$21 in Algorithm \ref{algo:blindChnlEst}. Here, Block-2 generates beamforming vectors $\mathbf{w}_{1}$ and $\mathbf{w}_{2}$ which are used for estimating $\tilde{\tau}_{1}$ and $\tilde{\tau}_{2}$ values, by another block that is excluded in Fig. \ref{est} (a). These values are fed as $\tilde{\tau}$ to the proposed OEB architecture, as shown in Fig. \ref{est} (a). It also shows that Block-2 has feedback information which is iteratively processed to generate the optimal beamforming vector $\bw_{opt}$. Eventually, the controller has been designed to feed various control signals to both Block-1 and Block-2 modules of the OEB architecture. These control signals enable OEB to set the systematic flow of information in every iteration of calculating the optimal beamforming vector. %the blind channel-estimation process.

\subsection{Micro-architectures of Block-1 \& Block-2}

\subsubsection{Block-1} 
\begin{figure}
	\centering
	\includegraphics[scale=0.480]{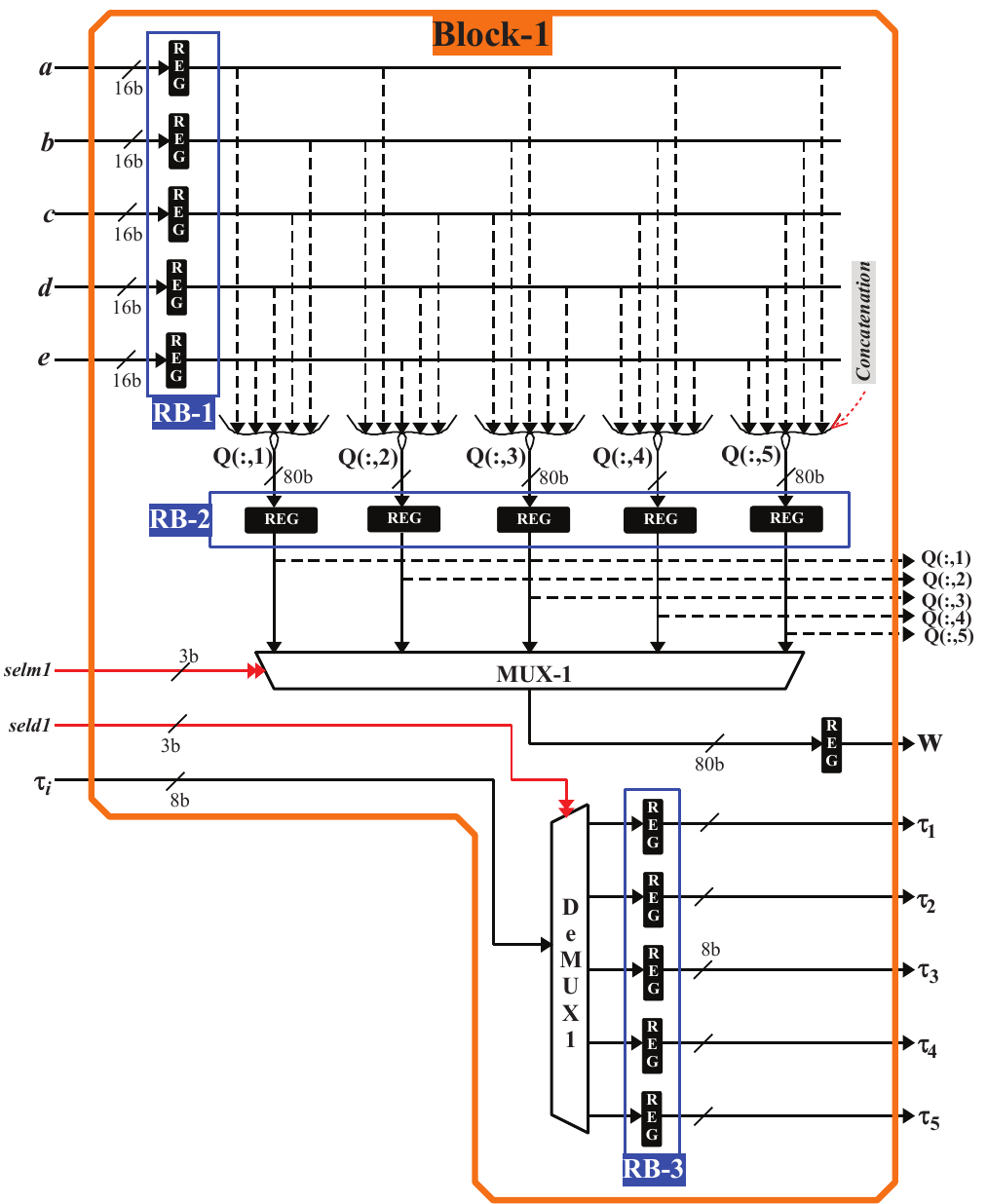}
	\caption{Proposed micro-architecture of the Block-1 submodule of OEB.}
	\label{block1}\
\end{figure}

A detailed explanation of the hardware micro-architectures of OEB sub-modules is presented here. Firstly, the design of Block-1 micro-architecture that comprises register banks (RBs) and steering logics is presented in Fig. \ref{block1}. It is fed (in single clock cycle) with five complex elements of $\mathbf{Q}$ matrix: \emph{a}, \emph{b}, \emph{c}, \emph{d}, and \emph{e} where each of them is 16-bit represented fixed-point value. They are the first row elements of $\mathbf{Q}$ matrix, referring to Fig. \ref{est} (b), and they are buffered in RB-1 of Block-1. Subsequently, RB-1 outputs are tapped and concatenated to generate five column vectors $-$$\mathbf{Q}$(:, 1) to $\mathbf{Q}$(:, 5)$-$ of 80 bit each which are stored in RB-2, as shown in Fig. \ref{block1}. Further, these values from RB-2 are tapped as parallel outputs that will be later fed to Block-2. On the other hand, RB-2 outputs are also routed via multiplexer MUX1 and are transmitted as $\mathbf{w}$ vector for the computation of $\tau_{i}$ $\forall$ \emph{i} = \{1, 2, 3, 4, 5\} values, as shown in Fig. \ref{block1}. It also shows that the computed $\tau_{i}$ real-values (each of 8 bit) are sequentially fed to Block-1 architecture where they are routed to RB-3 via de-multiplexer DeMUX1. Eventually, these five parallel $\tau_{i}$ values are also projected as outputs of Block-1 design.

\subsubsection{Block-2} 
\begin{figure}
	\centering
	\includegraphics[scale=0.440]{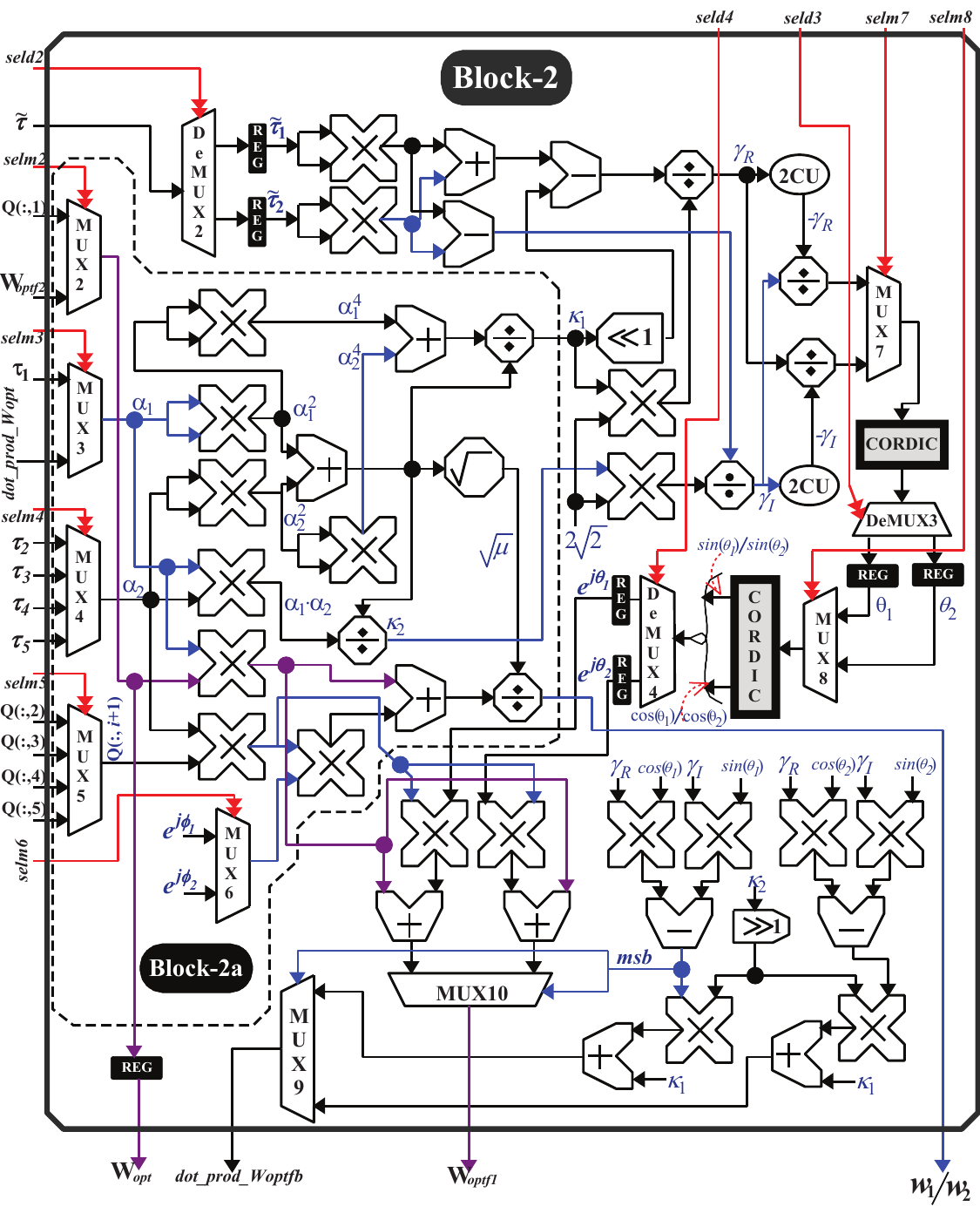}
	\caption{Proposed micro-architecture of the Block-2 submodule of OEB.}
	\label{block2}\
\end{figure}

The proposed micro-architecture of Block-2 shown in Fig. \ref{block2} has been designed to perform the operations of steps 8$-$21 in Algorithm \ref{algo:blindChnlEst}. To begin with, sub-module Block-2a executes steps 8$-$14 of Algorithm \ref{algo:blindChnlEst}, as shown in Fig. \ref{block2}. Here, MUX3 and MUX4 of Block-2a are responsible for assigning the input $\tau_{i}$ values (released from Block-1) to $\alpha_{1}$ and $\alpha_{2}$, respectively, corresponding to step 8 of Algorithm \ref{algo:blindChnlEst}. Similarly, MUX2 and MUX5 are used for routing $\mathbf{Q}$(:, 1) to $\mathbf{Q}$(:, 5) input vectors into the datapath, as shown in Fig. \ref{block2}. Subsequently, outputs of these multiplexers in Block-2a are processed by multipliers, adders, dividers, square-root computation unit, and MUX6, to generate $\kappa_{1}$, $\kappa_{2}$, and beamforming vectors ($\mathbf{w}_{1}$ and $\mathbf{w}_{2}$) that completes the operations till step 14 in Algorithm \ref{algo:blindChnlEst}. These $\mathbf{w}_{1}$ and $\mathbf{w}_{2}$ vectors are transmitted for computing the $\tilde{\tau}_{1}$ and $\tilde{\tau}_{2}$ values, respectively. They are fed as $\tilde{\tau}$ input to Block-2a where the de-multiplexer DeMUX2 routes and buffers such sequentially generated values as $\tilde{\tau}_{1}$ and $\tilde{\tau}_{2}$, as shown in Fig. \ref{block2}. It further shows that such $\tilde{\tau}$ and $\kappa$ values are arithmetically processed to compute $\gamma_{R}$ and $\gamma_{I}$ entities of step 15 in Algorithm \ref{algo:blindChnlEst}. These $\gamma$ values are passed through 2's complement units (2CUs) to realize ${-}\gamma_{R}$ and ${-}\gamma_{I}$ which are further divided and routed to coordinate rotations digital-computer (CORDIC) unit \cite{cordic} via multiplexer MUX7, referring Fig. \ref{block2}. Such MUX7 sequentially provides $\left(-\gamma_{I}/\gamma_{R}\right)$ and $\left(\gamma_{I}/-\gamma_{R}\right)$ values to the CORDIC unit that computes $\theta_{1}$ and $\theta_{2}$ values in resource shared manner. Subsequently, outputs of the CORDIC unit are de-multiplexed (by DeMUX3) to buffer $\theta_{1}$ and $\theta_{2}$ values that realizes step 16 of Algorithm \ref{algo:blindChnlEst}.

Another set of CORDIC unit with multiplexer MUX8 has been used for computing sin($\theta_{1}$), sin($\theta_{2}$), cos($\theta_{1}$), and cos($\theta_{2}$), as shown in Fig. \ref{block2}. These transcendental values are concatenated to realize $e^{(j{\cdot}\theta_{1})}$ and $e^{(j{\cdot}\theta_{2})}$ which are used for computing the $\mathbf{w}_{opt}$ values of steps 18 and 20 in Algorithm \ref{algo:blindChnlEst}, and they are fed to multiplexer MUX10. In parallel, $\gamma_{R}{\cdot}$cos($\theta_{1}$)$-$$\gamma_{I}{\cdot}$sin($\theta_{1}$) value has been generated and its MSB is used as select line of MUX10 whose output (denoted as $\mathbf{w}_{optf1}$) is the compared value between $\mathbf{w}_{opt}$ values of steps 17 to 21 in Algorithm \ref{algo:blindChnlEst}. Similarly, ${dot}\_{prod}\_{\mathbf{w}_{opt}}$ values of steps 18 and 20 in Algorithm \ref{algo:blindChnlEst} are computed and compared via multiplexer MUX9 whose output is denoted as ${dot}\_{prod}\_{\mathbf{w}_{optfb}}$ in Fig. \ref{block2}. It shows that the aforementioned outputs from MUX9 and MUX10 are buffered and fed back to MUX3 and MUX2, respectively, in order to run multiple iterations between steps 7 to 22 in Algorithm \ref{algo:blindChnlEst}.

\subsubsection{Controller} Controller of the proposed OEB generates various select-line values as control signals for multiplexers (i.e. \emph{selm1}$-$\emph{selm8}) and de-multiplexers (i.e. \emph{seld1}$-$\emph{seld4}) to operate multiple iterations and direct the data for valid operation, as shown in Fig. \ref{est}. As presented in Algorithm \ref{algo:blindChnlEst}, steps 2$-$3 and steps 11$-$14 operations wait for $\tau$ and $\tilde{\tau}$ values before transmitting the new $\mathbf{w}$ vector. Thereby, ${Ack}\_{Sig}$ has been used as a handshake/acknowledgement signal between such modules and the proposed OEB architecture. This ${Ack}\_{Sig}$ is fed to the controller of OEB that generates required control signals for Block-1 and Block-2 submodules. After all the iterations are executed, the optimal beamforming vector $\mathbf{w}_{opt}$ is eventually tapped from MUX2 as an output of the proposed OEB architecture, as shown in Fig. \ref{block2} and \ref{est}.

\subsection{{FPGA Implementation and ASIC Design Results}}

The proposed OEB architecture for finding the optimal beamforming vector has been implemented on a field-programmable gate-array (FPGA) platform. Static timing analysis of this design indicates that our OEB prototype can operate with the maximum clock frequency of 44.25 MHz and its critical path delay is 22.6 ns. Hardware utilization of the OEB architecture when implemented on Nexys4 DDR (xc7a100tcsg324-1) FPGA-board has been comprehensively presented in Table \ref{hard1}. It also shows the segregated hardware-consumption details of Block-1, Block-2, and controller micro-architectures of OEB. Here, Block-2 consumes the maximum FPGA resources of 3661 look-up tables (LUTs), 897 flip-flops (FFs), and 24 digital signal processor (DSP) cores.

Furthermore, the suggested OEB architecture has been ASIC synthesized and post-layout simulated in a 90 nm-CMOS technology node. Industry-standard electronics design automation (EDA) tools are used to perform frontend and physical design processes where the former includes functional validation, gate-level synthesis, static timing analysis, and post-synthesis simulations. On the other hand, the physical design process includes floorplan, standard-cell placement, power planning, signal routing, clock tree synthesis, timing verification, and finally the post-layout simulation. Outcomes of such ASIC design processes are presented in Table II where the design parameters of Block-1, Block-2, and Controller modules of the proposed OEB are separately listed. In addition, Table II also shows that an overall OEB design occupies an area of 0.21 mm$^2$ and it is capable of operating with a maximum clock frequency of 120 MHz. At the supply voltage of 1.2 V, it consumes dynamic power of 44385.7 pW while operating at 120 MHz of clock frequency and static power consumption of 816.61 pW, as shown in Table II.

%%%%%%%%%%%%%%%%%%%%%%%%%%%%%%%%%%%%%%%----------------------------------FPGA Results-------------------------------------------
\begin{table}[ht]
	\caption{Hardware Utilization of the Proposed OEB Architecture Implemented on Nexys4-DDR (xc7a100tcsg324-1) FPGA-Board}
	\centering
	\scalebox{1.0}{
		\begin{tabular}{ p{3 cm} || c | c | c || c  }
			\hline \hline
			& Block-1 & Block-2 & Controller & \textbf{OEB}\\
			\hline\hline
			Number of LUTs & 169 & 3661 & 150 & \textbf{3980}\\
			\hline
			Number of FFs & 371 & 897 & 43 & \textbf{1311}\\
			\hline
			Number of F7 MUXes & 80 & 0 & 0 & \textbf{80}\\
			\hline
			Number of DSP Cores & 0 & 24 & 0 & \textbf{24}\\
			\hline \hline
			
		\end{tabular}
	}
	\label{hard1}
\end{table}

%%%%%%%%%%%%%%%%%%%%%%%%%%%%%%%%%%%%%%%------------------ASIC Results-----------------------------------------------------------
\begin{table}[ht]
	\caption{{ASIC Synthesis and Post-Layout Simulation Results of Suggested OEB Architecture in UMC 90 nm-CMOS Process}}
	\centering
	\scalebox{1.0}{
		\begin{tabular}{ p{2.6 cm} || c | c | c || c  }
			\hline \hline
			& Block-1 & Block-2 & Controller & \textbf{OEB}\\
			\hline\hline
			Technology (nm) & 90 & 90 & 90 & \textbf{90}\\
			\hline
			Supply Voltage (V) & 1.2 & 1.2 & 1.2 & \textbf{1.2}\\
			\hline
			Silicon Area (mm$^{2}$) & 0.012 & 0.173 & 0.00033 & \textbf{0.21}\\
			\hline
			Standard Cell Count & 946 & 17208 & 39 & \textbf{18193}\\
			\hline
			Crit. Path Delay (ps) & 1744 & 8784 & 1411 & \textbf{8784}\\
			\hline
			Max. Clk Freq. (GHz) & 0.57 & 0.114 & 0.71 & \textbf{0.12}\\
			\hline
			Dynamic Power (pW) & 2285.2 & 42060.5 & 39.99 & \textbf{44385.7}\\
			\hline
			Leakage Power (pW) & 46.6 & 768.7 & 1.311 & \textbf{816.61}\\
			\hline
			Bit Quantization (bits) & 16 & 16 & 16 & \textbf{16}\\
			\hline \hline
			
		\end{tabular}
	}
	\label{hard2}
\end{table}

\section{Numerical results}\label{sec:numres}
In this section, through numerical computation we evaluate the performance of the proposed algorithm by varying relevant system parameters. As the article focuses on improving the energy delivery, as elaborated in section \ref{sec:sysmod}, we consider only those parameters that are relevant to energy transmission and harvesting. Table \ref{tab:param} summarises the parameters considered in the evaluation of the proposed algorithm, where the values of some parameters are taken from \cite{LiuTC2014, zengTCOM2014, sarmaAccess2019}. %\txtred{\cite{abeywickramaTSP2018,zengTCOM2014,sarmaAccess2019}}.

\begin{table}[!h]
	\centering
	\caption{Table of the parameters used in numerical computation.}
	\begin{tabular}{c c}
		\hline
		Parameter & Value(s)/range \\
		\hline
		Transmit power, $P_t$ & $3 - 10$ Watts \\
		No. of antennas at the Tx, $N$ & $5, 10$ \\
		Antenna gain, $G$ & $1$\\
		Distance between Energy Tx \& Rx, $L$ & $3 - 10$ meters \\
		Pathloss exponent, $\beta$ & $3$ \\
		Rician factor, $K_f$ & $2, 10$ \\
		Resistance (Energy storage), $R$ &  $100$ $\Omega$ \\
		Capacitance (Energy storage), $C$ & $1$mF \\
		Initial charge level, $q_0$ & $1.5$ mC \\
		Maximum charge level, $q_m$ & $3$ mC \\
		Time limit, $t_{lim}$ & $100$ seconds\\
		\hline  
	\end{tabular}
\label{tab:param}
\end{table}  

As the maximum distance between the energy transmitter and the energy receiver is $10$ meters (cf. Table \ref{tab:param}), we assume the channel is characterised by a dominant \cmnt{presence of a significant} line-of-sight (LoS) component. Similar to \cite{zengTCOM2014}, we consider Rician fading channel to model the channel between the two devices. Specifically, we consider the following three channels for the performance evaluation (i) Fading channel with Rician factor ($K_f$) = 2, (ii) Fading channel with $K_f= 10$ and (iii) Deterministic channel (corresponding to $K_f \to \infty$). We also incorporate the path-loss with the help of the path-loss exponent $\beta$. It should be noted that the plots presented in this section are generated by averaging over 1000 realisations of fading channels mentioned previously. Furthermore, we assume the following piece-wise linear energy harvesting efficiency function for calculating the harvested power at the receiver. 
\begin{align}\label{eq:plEHSim}
\eta = \begin{cases}
0, & P_r < p_{th,0}  \\
\eta_0\frac{(Pr - p_{th,0})}{p_{th,1} - p_{th,0}}, & p_{th,0} \le P < p_{th,1}\\
\eta_0 + (\eta_1 - \eta_0)\frac{(P_r - p_{th,1})}{p_{th,2} - p_{th,1}},  & p_{th,1} \le P_r < p_{th,2} \\
\eta_1 + (\eta_2 - \eta_1)\frac{(P_r - p_{th,2})}{p_{th,3} - p_{th,2}}, & p_{th,2} \le P_r \le p_{th,3} \\
\eta_2, & P_r > p_{th,3}, 
\end{cases}
\end{align}
with $\eta_0 = 0.4$, $\eta_1 = 0.6$, $\eta_2 = 0.65$, $p_{th,0} = 10^{-6}$ , $p_{th,1} = 10^{-5}$, $p_{th,2} = 10^{-4}$ and $p_{th,3} = 10^{-3}$. 
%\subsubsection*{Simulation of the channel gains} 
\begin{figure*}[!h]
	\centerline{\subfigure[]{\includegraphics[scale=0.6]{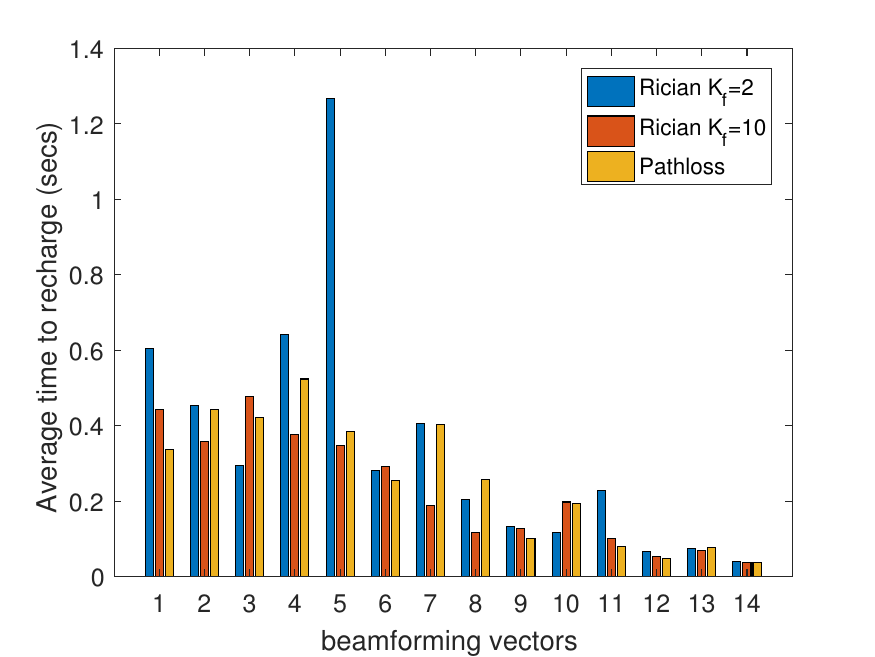}
	%\caption{Average time-to-recharge due to different beamforming vectors during the feedback acquisition and processing phase for considered channels. The number of antennas at the transmitter is $5$. The transmit power is set to $10$ Watts, and the distance between the energy transmitter and harvester is $5$ meters.}
	\label{fig:avgTTR_bmfrmVec}}
     \hfil
     \subfigure[]{\includegraphics[scale=0.6]{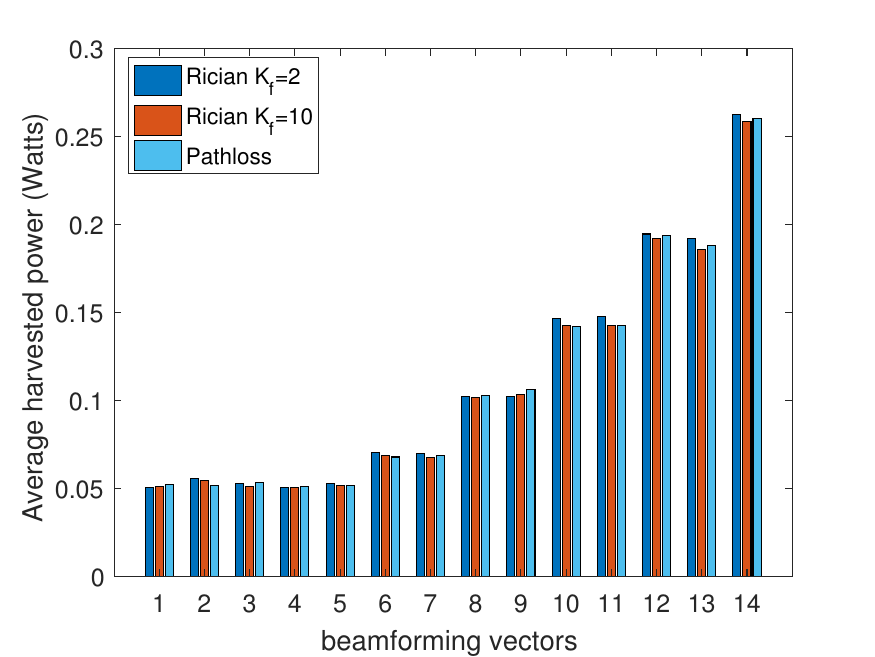}
     \label{fig:avgHrvPwr_bmfrmVec}}}
    \caption{Average time-to-recharge and average harvested power due to different beamforming vectors during the feedback acquisition and processing phase for considered channels. The transmit power is set to $10$ Watts and the distance between the energy transmitter and harvester is $5$ meters.}  
   \label{fig:ttrHrvPower5ant}
\end{figure*}

In Fig. \ref{fig:ttrHrvPower5ant}, we depict the average time-to-recharge and average harvested power due to beamforming vectors employed by the proposed algorithm during the FAP phase. As the energy transmitter considered for this figure has $5$ antennas, the FAP phase spans time-to-recharge duration of $13$ beamforming vectors indexed by ordinal numbers. The first five set of bars in Figs. \ref{fig:avgTTR_bmfrmVec} and \ref{fig:avgHrvPwr_bmfrmVec} correspond to five beamforming vectors obtained from columns of $\mb{Q}$. The following eight set of bars correspond to the intermediate beamforming vectors obtained through linear combinations of columns of $\mb{Q}$. The last set of bars correspond to the optimal beamforming vector obtained at the end of the FAP phase. It is evident that the sixth beamforming vector onward the harvested power increases and consequently, the time-to-recharge decreases as the algorithm produces vectors that approach the optimal beamforming vector. It should be noted that time-to-recharge values were capped at $t_{lim}$ mentioned in Table \ref{tab:param}. 
%\begin{figure}[!h]
%	\centering
%	\includegraphics[scale=0.6]{Figures/barChartAvgHrvPwr5ant_1k}
%	\caption{Average harvested power due to different beamforming vectors during the feedback acquisition and processing phase for considered channels. The transmit power is set to $10$ Watts and the distance between the energy transmitter and harvester is $5$ meters.}
%	\label{fig:avgHrvPwr_bmfrmVec}
%\end{figure}

\begin{figure}[!h]
	\centering
	\includegraphics[scale=0.6]{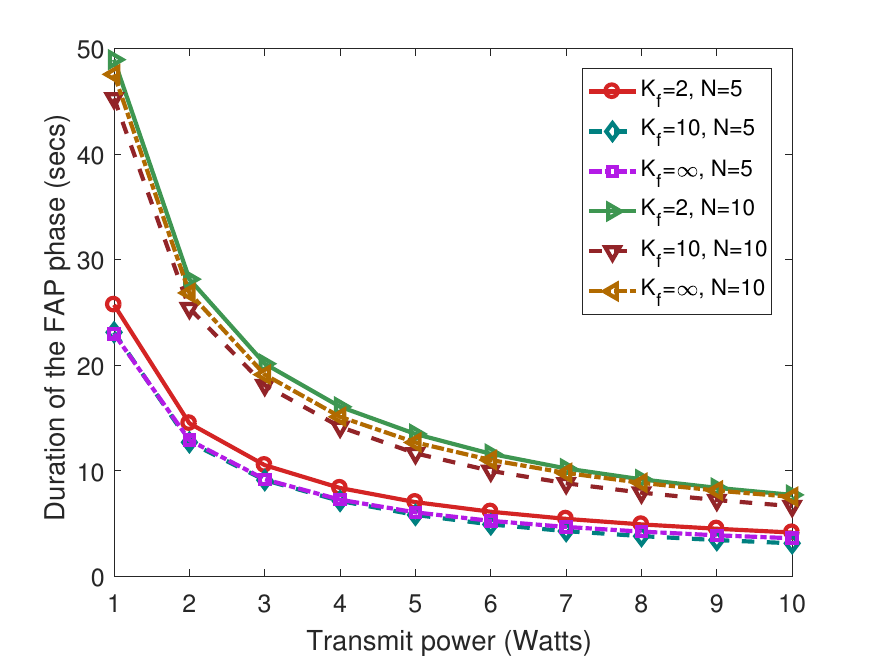}
	\caption{Average duration of the FAP phase with respect to the transmit power for different channels types and $5, 10$ antennas at the transmitter. The distance between the energy transmitter and harvester is $5$ meters.}
	\label{fig:avgEstPhsDurVsTxPower}
\end{figure}

\begin{figure}[!h]
	\centering
	\includegraphics[scale=0.6]{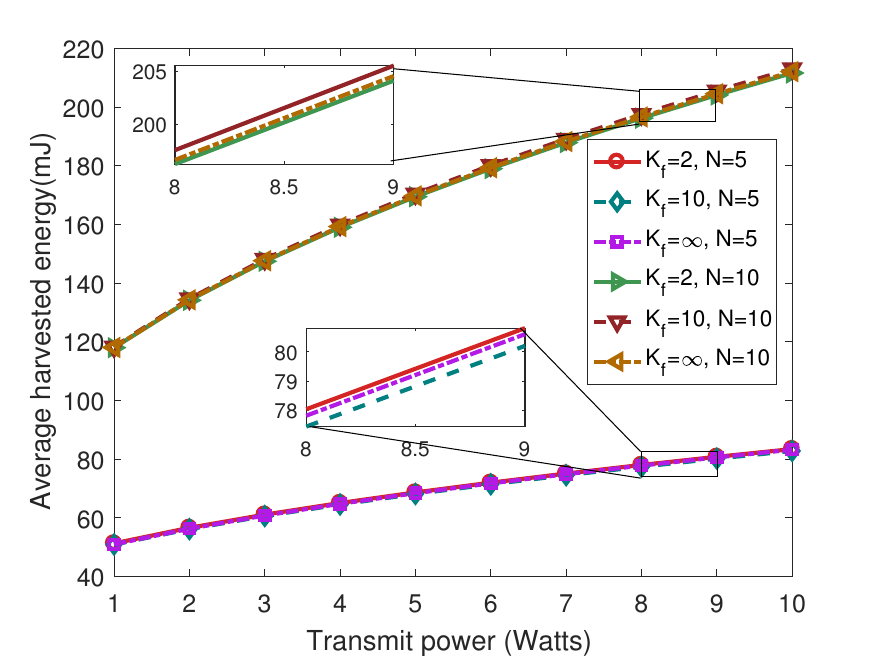}
	\caption{Average harvested energy during the FAP phase vs. transmit power for different channel types and $5, 10$ antennas at the transmitter. The distance between the energy transmitter and harvester is $5$ meters.}	
	\label{fig:avgHrvPwrVsTxPower}
	\end{figure}

Fig. \ref{fig:avgEstPhsDurVsTxPower} plots the duration of the FAP phase with respect to the transmit power at the energy transmitter for $5$ and $10$ antenna systems when the receiver is $5$ meters away. The duration of the FAP phase of the latter is almost double of the former and can be explained by counting the number of slots in the respective FAP phases (i.e., $13$ and $28$). While a longer FAP phase might seem disadvantageous, one must not forget that a proportionately larger number of recharging cycle (and the subsequent information transmission) is completed during that phase. The duration of the FAP phase for both $5$ and $10$ antenna transmitters decreases with the transmit power, which is intuitive. Fig. \ref{fig:avgHrvPwrVsTxPower}, on the other hand, portrays the variation of the harvested energy with respect to the transmit power. The $10$-antenna transmitter is able to deliver approx $3\times$ energy compared with the $5$-antenna transmitter due to $2\times$ number antennas and $\approx2\times$ slots in the FAP phase. Similar to the Fig. \ref{fig:avgEstPhsDurVsTxPower}, the increase in the harvested energy with transmit power is apparent.     

%\begin{figure}[!h]
%	\centering
%	\includegraphics[scale=0.6]{Figures/totEstRchrgTimeVsDist_5_10ant1}
%	\caption{Fig. \ref{fig:avgEstPhsDurVsTxPower} and \ref{fig:avgHrvPwrVsTxPower} together.}
%	\label{fig:estPhaseHrvPwrVsTxPower}
%\end{figure}

\begin{figure}[!h]
	\centering
	\includegraphics[scale=0.6]{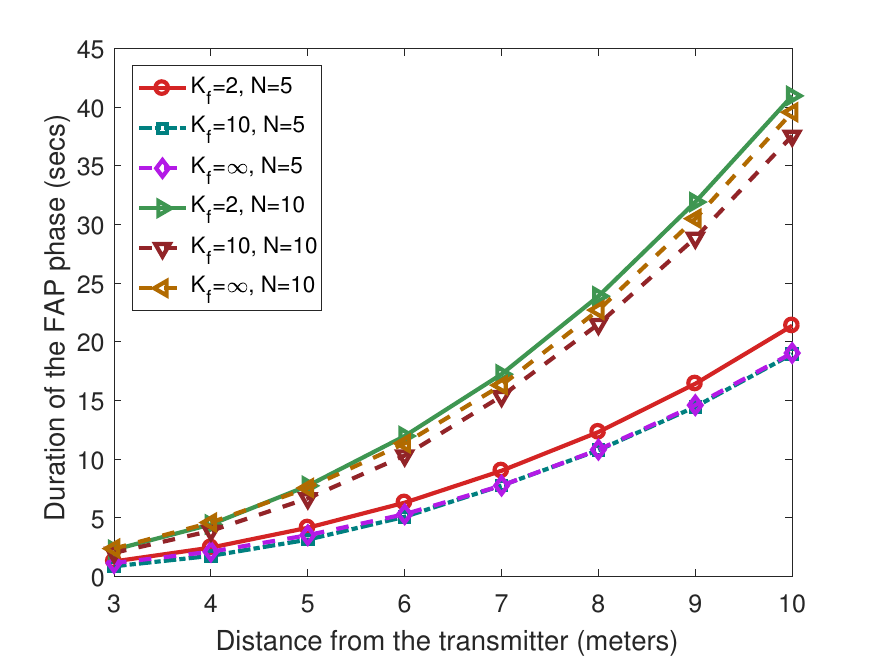}
	\caption{Average duration of the FAP phase with respect to the distance between the energy transmitter and harvester for different channel types and $5, 10$ antennas at the transmitter. The transmit power is set to $10$ W.}
	\label{fig:estPhaseDurVsTxDist}
\end{figure}

\begin{figure}[!h]
	\centering
	\includegraphics[scale=0.6]{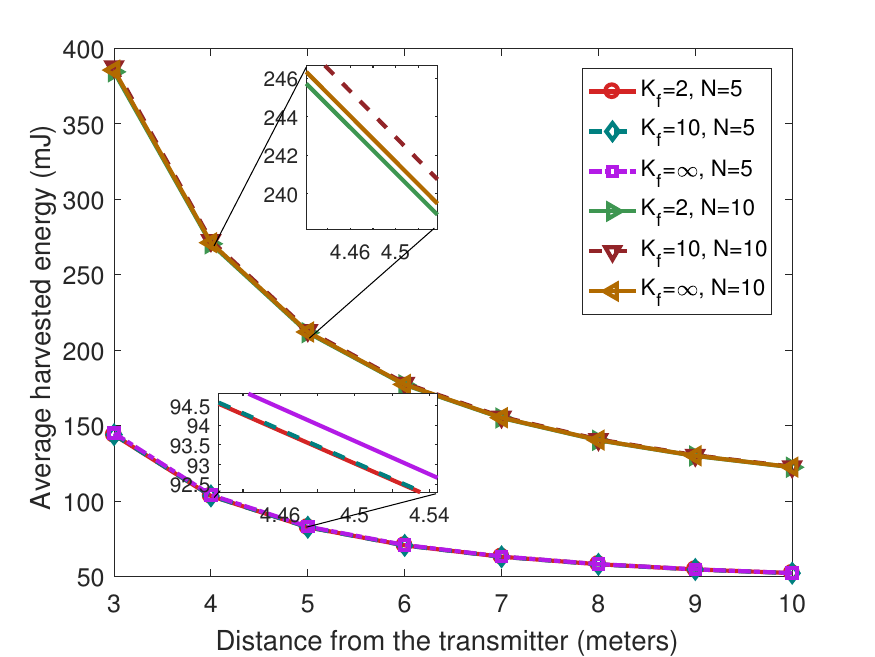}
	\caption{Average harvested energy during the FAP phase with respect to the distance between the energy transmitter and harvester for different channel types and $5, 10$ antennas at the transmitter. The transmit power is set to $10$ W.}
	\label{fig:avgHrvPwrVsTxDist}
\end{figure}
In Fig. \ref{fig:estPhaseDurVsTxDist} and \ref{fig:avgHrvPwrVsTxDist}, we plot the duration of the FAP phase and the energy harvested in that phase, respectively, with respect to the distance of the receiver from the transmitter while fixing the transmit power to $10$ Watts. As the distance increases, the received power decreases and the subsequently the duration of the FAP phase increases for both $5$ and $10$ antenna transmitters. In both plots, the values plotted in $y$-axis varies in proportion to the number of antennas, similar to that of Fig. \ref{fig:avgEstPhsDurVsTxPower} and \ref{fig:avgHrvPwrVsTxPower}.
%%%\begin{figure}[!h]
%%%	\label{fig:estPhaseHrvPwrVsTxDist}
%%%	\centering
%%%	\includegraphics[scale=0.6]{Figures/estPhaseHrvPwrVsTxDist_1k}
%%%	\caption{Fig. \ref{fig:estPhaseDurVsTxDist} and \ref{fig:avgHrvPwrVsTxDist} in the same figure.}
%%%\end{figure}

%\begin{figure}[!h]
%	\centering
%\end{figure}
%Number of antennas vs time taken to complete estimation process
%
%Distance vs time to complete estimation
%
%Transmit power vs time to complete estimation
\section{Conclusion}\label{sec:cnclsn}
In this article, we presented a scheme to find the optimal beamforming vector for the energy beamforming at a transmitter from indirect feedback collected from a energy receiver. The proposed scheme exploits the dynamics of the harvest-then-transmit protocol and the inherent latency of the charging process to extract a temporal feedback about the beamforming vector used by the transmitter. As the receiver's energy is preserved in the proposed scheme, a higher throughput or reduced latency in information transmission, and energy saving at the transmitter are guaranteed, leading to an improved WPT system. The presented scheme is supplemented by an algorithm and its hardware architecture for facilitating its adoption in various real-world WPT implementations. Hardware utilisation details, ASIC synthesis and post-layout simulation results were presented to substantiate the proposed architecture. Lastly, through numerical results we demonstrated the performance of the proposed scheme with respect to various system parameters.   

%\txtblu{The current art}
\bibliographystyle{IEEEtran}
\bibliography{IEEEabrv,wehChnlEst}	
\end{document}